\documentclass[12pt]{article}

\usepackage{amsfonts,amssymb,amsmath}
\usepackage{color}
\usepackage{theorem,cite}

\numberwithin{equation}{section}
\newcommand{\nonu}{\nonumber \\}

\newcommand{\mb}[1]{\quad\mbox{#1}\quad}
\newcommand{\beq}{\begin{equation}}
\newcommand{\eeq}{\end{equation}}
\newcommand{\bea}{\begin{eqnarray}}
\newcommand{\eea}{\end{eqnarray}}
\newcommand{\beano}{\begin{eqnarray*}}
\newcommand{\eeano}{\end{eqnarray*}}

\definecolor{brique}{rgb}{.9,.2,0}
\definecolor{blvert}{rgb}{0,.8,.85}
\definecolor{vertcl}{rgb}{0,1,.7}
\newcommand\vertcl[1]{\textcolor{vertcl}{#1}}
\newcommand\blvert[1]{\textcolor{blvert}{#1}}
\newcommand\brique[1]{\textcolor{brique}{#1}}
\def\lapth{
\begin{picture}(164,70)(0,-15)\thicklines
\put(0,0){\vertcl{\rule{20pt}{4pt}}}
\put(19,1){\vertcl{\line(1,3){23}}} 
\put(20,1){\vertcl{\line(1,3){23}}} 
\put(21,1){\vertcl{\line(1,3){23}}}
\put(22,1){\vertcl{\line(1,3){23}}}
\put(45,70){\vertcl{\line(1,-3){23}}} 
\put(44,70){\vertcl{\line(1,-3){23}}} 
\put(43,70){\vertcl{\line(1,-3){23}}}
\put(42,70){\vertcl{\line(1,-3){23}}}
\put(2,24){\vertcl{\rule{120pt}{4pt}}}
\put(65,0){\vertcl{\rule{60pt}{4pt}}}
\put(5,37){\Huge{\brique{\textbf{L}}}} 
\put(62,37){\Huge{\brique{\textbf{PTh}}}}
\put(12,-8){\blvert{\rule{92pt}{3.5pt}}}
\put(24,-15){\blvert{\rule{57pt}{3.5pt}}}
\put(36,-22){\blvert{\rule{30pt}{3.5pt}}}
\end{picture}
\raisebox{35pt}{
\begin{minipage}{320pt}\begin{center}
\textbf{Laboratoire d'Annecy-leVieux de Physique
Th\'eorique}\\[4ex]
website: \texttt{http://lappweb.in2p3.fr/lapth-2005/}
\end{center}
\end{minipage}}\\
\vspace{10pt}\quad \hrulefill\\
\vspace{10pt}}

\def\cG{{\cal G}}    \def\cH{{\cal H}}    
        
    \def\cN{{\cal N}}    \def\cO{{\cal O}}
\def\cP{{\cal P}}        
        
\def\cV{{\cal V}}

\def\cNb{\overline\cN}

\def\fS{{\mathfrak S}}

\def\fm{{\mathfrak m}}
\def\fn{{\mathfrak n}}
\def\fp{{\mathfrak p}}

\newcommand{\bA}{{\mathbb A}}

\newcommand{\CC}{{\mathbb C}}
\newcommand{\EE}{{\mathbb E}}
\newcommand{\II}{{\mathbb I}}

\newcommand{\ZZ}{{\mathbb Z}}

\newcommand{\wt}[1]{\widetilde{#1}}
\newcommand{\wb}[1]{\overline{#1}}

 \newcommand{\spn}{\text{span}}

\newcommand{\atopn}[2]{\genfrac{}{}{0pt}{}{#1}{#2}}

\begin{document}

%%%%%%%%%%%%%%%%%%%%%%%%%%%%%%%%
%%%%%  HEADINGS POUR DRAFT  %%%%%%

\markright{\today\dotfill DRAFT\dotfill }
\pagestyle{myheadings}
 \pagestyle{empty}
\setcounter{page}{0}

%%%\section{Title}
\hspace{-1cm}\lapth

\vfill

\begin{center}

{\LARGE{\sffamily Bethe equations for generalized Hubbard 
models}}\\[1cm]

{\large V. Fomin, L. Frappat
and E. Ragoucy\footnote{
fomin@lapp.in2p3.fr, frappat@lapp.in2p3.fr, 
ragoucy@lapp.in2p3.fr}\\[.484cm] 
\textit{Laboratoire de Physique Th{\'e}orique LAPTH\\[.242cm]
Universit{\'e} de Savoie -- CNRS (UMR 5108) \\[.242cm]
9 chemin de Bellevue, BP 110, F-74941  Annecy-le-Vieux Cedex, France. 
}}
\end{center}
\vfill

\begin{abstract}
We compute the eigenfunctions, energies and Bethe equations for 
a class of generalized integrable Hubbard models based on 
$gl(\fn|\fm)\oplus gl(2)$ superalgebras. The Bethe equations appear 
to be similar to the Hubbard model ones, up to a phase due to the 
integration of a subset of `simple' Bethe equations. 
We discuss relations with 
AdS/CFT correspondence, and with condensed matter physics.
\end{abstract}

\vfill
% \rightline{\tt arXiv: [hep-th]}
\rightline{LAPTH-1334/09}
\rightline{June 2009}

\newpage
\pagestyle{plain}

\section{Introduction}

The Hubbard model was first introduced as a model to describe the effects
of the strongly correlated d-electrons in transition metals
\cite{Hubbard,Gutzwiller}. It was shown that this model possesses a Mott
metal-insulator phase transition \cite{Mott,Hubbard3} and is relevant to
the studies of high-$T_{c}$ superconductivity \cite{Anderson,Affleck}. Few
exact results are known for the two or three-dimensional Hubbard model but
nevertheless these models are still actively investigated.

In contrast, the one-dimensional Hubbard model is integrable and was first
diagonalized by means of coordinate Bethe Ansatz by Lieb and Wu
\cite{LiebWu}. However its integrable structure is rather complicated in
comparison with the usual formalism of the spin chains (for a review see
e.g. \cite{book} and references therein). The Hubbard model $R$-matrix was
first introduced by Shastry \cite{shastry} and Olmedilla et al
\cite{akutsu} (by coupling two XX model $R$-matrices with U-interaction
term). The proof of the Yang--Baxter relation was given by Shiroishi and
Wadati \cite{shiro}. A lot of extended Hamiltonians were also proposed to
connect the model with high-$T_{c}$ superconductivity effects
\cite{EKScomp,bariev,angela} (and references therein).

Generalization of one-dimensional Hubbard model to $gl(\fn)$ case via
$R$-matrix was made by Maassarani et al \cite{maassa}. Extension of
Maassarani's approach to superalgebras $gl(\fn|\fm)$ was given in
\cite{XX}, and a "universal" presentation (including a generalization of
Maassarani's approach) was proposed in \cite{FFR2009}. It is based on the
decomposition of an arbitrary vector space into a direct sum of two
subspaces, the two corresponding orthogonal projectors allowing one to
define an $R$-matrix of a universal XX model, and then of a Hubbard model
using a Shastry type construction. The construction is very general, since
the two XX models that are used can be based on different (super)algebras.
The QISM approach ensures the integrability of the models, leading to local
Hubbard-like Hamiltonians thanks to the appropriate behavior of the
obtained $R$-matrices.

Recent studies of the Hubbard model and its generalizations, were motivated
by its "recent" applications in $N=4$ super Yang--Mills theory (SYM) -- for
example, see \cite{Rej:2005qt} and references therein. Although it appeared
that the one-dimensional Hubbard model is not the proper answer due to
transcendental contributions to the anomalous dimensions (in the $su(2)$
subsector of the theory), one may find new directions in this field by
studying integrable extensions of the one-dimensional Hubbard model.
Another aspect lies in the possibility of applications to condensed matter
physics, particularly when dealing with small rank algebras.

In this paper, we focus on a subcase of the "universal" presentation
where the superalgebra for one XX model is 
$gl(\fn|\fm)$ while the other XX model is based on the algebra 
$gl(2)$, the projectors for both XX models being one-dimensional. This 
kind of model is simple enough to investigate further the problem of  
Bethe Ansatz Equations (see \cite{FFR2009}), and presents interesting 
features to find some applications in condensed matter physics as well 
as some relevence in AdS/CFT correspondence.

The paper is organised as follows. In section \ref{sec:genHub} we 
prepare the notations
and recall some results on generalized Hubbard model based on unitary
(super)algebras. In section \ref{sec:gl21}, we solve via the 
coordinate Bethe ansatz a
toy model constructed on $gl(2|1)\oplus gl(2)$ with one-dimensional 
projectors and we generalize the
obtained results to $gl(\fn|\fm)\oplus gl(2)$ models in  section 
\ref{sec:glmn}. In section \ref{sec:gl22gl2 pi particle}, 
we study a model based on 
$gl(2|2)\oplus gl(2)$, where the projector for the $gl(2|2)$ part is now 
two-dimensional. We argue on the relevence of this model for the 
AdS/CFT correspondence. In section \ref{sec:JordWig}, we perform the 
Jordan--Wigner transformation for $gl(2|1) \oplus gl(2)$ and $gl(2|2) \oplus 
gl(2)$ models to construct physical Hamiltonians.

\section{Some results on $gl(\fn|\fm)$ Hubbard model\label{sec:genHub}}

We remind in this section the necessary notations for self-consistency and
understanding of the paper. For more details, we refer the reader to 
\cite{FFR2009}.

The starting point is the definition of the $R$-matrix of an XX model 
based 
on the superalgebra $\cG=gl(\fn|\fm)$. Consider the graded vector 
space 
$\cV = 
\CC^{\fn|\fm}$ (with possibly $\fm=0$) and define the $\ZZ_2$-grading 
on 
indices $j$ by 
\beq
[j]=\begin{cases} 0 \mb{for}1 \le j \le \fn\\
1 \mb{for} \fn+1 \le j \le \fn+\fm.
\end{cases}\label{eq:grading}
\eeq
 Accordingly, the elementary matrices $E^{ij}$ (with entry 1 at row 
$i$ 
and column $j$ and 0 elsewhere) have grade $[E^{ij}] = [i]+[j]$.
We introduce a set of integers $\cN \subset \{1,\ldots,\fn+\fm\}$ and 
$\cNb$ 
its 
complementary subset in $\{1,\ldots,\fn+\fm\}$. We then define the 
following 
projectors:
\begin{eqnarray}
\pi = \sum_{j\in\cN} E^{jj} \quad,\quad 
\wb\pi = \pi^{(\cNb)} = \sum_{\bar{\jmath}\in\cNb} E^{\bar{\jmath}\bar{\jmath}}
\label{def:pi}
\end{eqnarray}
as well as (using auxiliary space notation)
\begin{eqnarray}
\Sigma_{12} &=& 
\pi_{1}\,\wb\pi_{2}+\wb\pi_{1}\,\pi_{2} \,.
\label{def:univSigma}
\end{eqnarray}
The $R$-matrix of a XX model based on $gl(\fn|\fm)$ with projectors 
$(\pi,\wb\pi)$ is given by:
\begin{equation}
R_{12}(\lambda) = \Sigma_{12}\,P_{12} + \Sigma_{12}\,\sin\lambda +
(\II\otimes\II-\Sigma_{12})\,P_{12}\,\cos\lambda
\label{def:univRXX}
\end{equation}
where $P_{12}$ is the graded permutation operator and $\lambda \in 
\CC$ 
the spectral parameter. It obeys Yang--Baxter equation, is unitary and 
regular.

\medskip

The $R$-matrix for (generalized) Hubbard models is obtained by 
coupling the
$R$-matrices $R^{\uparrow}_{12}(\lambda)$ and
$R^{\downarrow}_{12}(\lambda)$ of two independent XX models, the 
coupling
constant being related to the potential $U$ of the Hubbard model under
consideration \cite{shastry,maassa,book,FFR2009,XX}. We stress that 
the two
XX models can be based on two different (super)algebras 
$\cG_{\uparrow}$
and $\cG_{\downarrow}$, with two different (graded) vector spaces
$\cV_{\uparrow}$ and $\cV_{\downarrow}$ and two different projectors
$\pi_{\uparrow}$ and $\pi_{\downarrow}$, associated to two different 
sets
$\cN_{\uparrow}$ and $\cN_{\downarrow}$.

Introducing the parity matrix $C_\alpha$ ($\alpha = \uparrow$ or 
$\downarrow$):
\begin{equation}
C_\alpha = \sum_{j\in\cN} E_{\alpha}^{jj} - \sum_{\bar{\jmath}\in\cNb} 
E_{\alpha}^{\bar{\jmath}\bar{\jmath}} = 
\pi_\alpha-\wb\pi_\alpha\,,
\label{eq:matC}
\end{equation}
the $R$-matrix of the Hubbard model based on the pair of 
(super)algebras
$\cG_{\uparrow}$ and $\cG_{\downarrow}$ is given by:
\begin{equation}
R^{\uparrow\downarrow}_{12}(\lambda_{1},\lambda_{2}) =
R^{\uparrow}_{12}(\lambda_{12})\,R^{\downarrow}_{12}(\lambda_{12}) +
\frac{\sin(\lambda_{12})}{\sin(\lambda'_{12})} \,\tanh(h'_{12})\,
R^{\uparrow}_{12}(\lambda'_{12})\,C_{\uparrow 1}\,
R^{\downarrow}_{12}(\lambda'_{12})\,C_{\downarrow 1}
\label{def:R-XXfus}
\end{equation}
where $\lambda_{12}=\lambda_{1}-\lambda_{2}$ and
$\lambda'_{12}=\lambda_{1}+\lambda_{2}$. 
In the same way, $h'_{12}=h(\lambda_{1})+h(\lambda_{2})$ and the 
function $h(\lambda)$ is such that 
\begin{equation}
\sinh(2h)=U\, \sin(2\lambda) \,.
\label{eq:h-lambda}
\end{equation}
Note that the site 1 for 
$R^{\uparrow\downarrow}_{12}$ is composed from the tensor product of 
the site $"1\uparrow"$ appearing in the matrix $R^{\uparrow}_{12}$ by 
the site $"1\downarrow"$ which is in the matrix $R^{\downarrow}_{12}$. 
This is obviously the same for any site we will consider in the 
following.

The $R$-matrix (\ref{def:R-XXfus}) is symmetric, regular and 
satisfies the
unitary relation. Moreover, when the relation (\ref{eq:h-lambda}) 
holds,
the $R$-matrix (\ref{def:R-XXfus}) satisfies the Yang--Baxter 
equation:
\begin{eqnarray}
R^{\uparrow\downarrow}_{12}(\lambda_{1},\lambda_{2})\, 
R^{\uparrow\downarrow}_{13}(\lambda_{1},\lambda_{3})\, 
R^{\uparrow\downarrow}_{23}(\lambda_{2},\lambda_{3}) 
&=& 
R^{\uparrow\downarrow}_{23}(\lambda_{2},\lambda_{3})\, 
R^{\uparrow\downarrow}_{13}(\lambda_{1},\lambda_{3})\, 
R^{\uparrow\downarrow}_{12}(\lambda_{1},\lambda_{2})\,.
\end{eqnarray}
Being equipped with an $R$-matrix with all required properties, we can
proceed to define the corresponding quantum integrable system, by
performing the following steps: monodromy matrix, transfer matrix and
 Hamiltonian. 
The $L$-site monodromy matrix is given
\begin{equation}
T_{a<b_{1}\ldots b_{L}>}(\lambda) =
R^{\uparrow\downarrow}_{ab_{1}}(\lambda,0)\ldots R^{\uparrow\downarrow}_{ab_{L}}(\lambda,0)
\end{equation}
and its transfer matrix is the (super)trace in the auxiliary space:
\begin{equation}
t(\lambda)=tr_{a}T_{a<b_{1}\ldots b_{L}>}(\lambda)\,.
\end{equation}
Then the generalized Hubbard Hamiltonian reads
\begin{equation}
H = \frac{d}{d\lambda} \ln t(\lambda) \bigg\vert_{\lambda=0} 
= \sum_{x=1}^{L} H_{x,x+1}
\label{eq:HubHam}
\end{equation}
with 
\begin{equation}
H_{x,x+1} = (\Sigma P)_{\uparrow \; x,x+1 } + (\Sigma P)_{\downarrow 
\;x,x+1 } + u \, C_{\uparrow k} \, C_{\downarrow k} \,,
\label{eq:twositesham}
\end{equation}
where we have used periodic boundary conditions. The notation 
$\cO_{\uparrow\;x,x+1}$ means that the operator $\cO$ acts non 
trivially in the parts $x\uparrow$ and $(x+1)\uparrow$ only. It acts 
as identity on all the sites different from $x$ and $x+1$ and also on 
the parts $x\downarrow$ and $(x+1)\downarrow$ of sites $x$ and $x+1$.
Explicitly, one has
$$
(\Sigma P)_{\alpha\;x,x+1 } = \sum_{j, \bar{
\jmath}}\Big\{(-1)^{[\bar{\jmath}]}E^{j \bar{
\jmath}}_{\alpha\;x}E^{\bar{\jmath} j }_{\alpha\;x+1}  +
(-1)^{[{j}]} E^{\bar{\jmath} j}_{\alpha\;x}
E^{j \bar{\jmath}}_{\alpha\;x+1}\Big\}
$$
with $\alpha = \;\uparrow$ or $\downarrow$. The indices $j$ and
$\bar{\jmath}$ run over $\cN$ and $\cNb$ respectively.

One can also define the momentum operator:
\beq
\exp(i\hat\fp) = t(0) = P_{1L}\,P_{2L}\,\ldots P_{L-1,L}\,.
\eeq

\section{$gl(2|1) \oplus gl(2)$ model\label{sec:gl21}}
\subsection{Preliminaries}

We consider an example of the above model for a particular algebra
$gl(2|1)_{\uparrow} \oplus gl(2)_{\downarrow}$. This notation
means that for different spin value up or down we take different
algebras in the construction of the Hamiltonian. The generic
expression of the $L$-site Hamiltonian is still the same:
   \begin{equation}
H_{gl(2|1) \oplus gl(2)} = \sum_{x=1}^{L} \left[ (\Sigma P)_{\uparrow 
\;
x,x+1 } + (\Sigma P)_{\downarrow \; x,x+1 } \right] + u \sum_{x=1}^L 
\left(C_{\uparrow x} \, C_{\downarrow x} \right),
\label{gl21gl2 Ham}
    \end{equation}
but we now choose the projectors $\pi_\uparrow$ and $\pi_\downarrow$ 
to be 
such that
    \begin{eqnarray}
&& (\Sigma P)_{\uparrow \; x,x+1 } = E^{12}_{\uparrow \; x}
E^{21}_{\uparrow \; x+1} + E^{21}_{\uparrow \; x} E^{12}_{\uparrow \; 
x+1}
- E^{13}_{\uparrow \; x}E^{31}_{\uparrow \; x+1} + E^{31}_{\uparrow \;
x}E^{13}_{\uparrow \; x+1}\,, \\
&& (\Sigma P)_{\downarrow \; x,x+1 } = E^{12}_{\downarrow \; x}
E^{21}_{\downarrow \; x+1} + E^{21}_{\downarrow \; x}
E^{12}_{\downarrow \; x+1}\,, \\
&& C_{\uparrow\; x} = E^{11}_{\uparrow\; x} - E^{22}_{\uparrow\; x} -
E^{33}_{\uparrow\; x} \mb{,}
 C_{\downarrow\; x} = E^{11}_{\downarrow\; x} - 
E^{22}_{\downarrow\; x}\,.
    \end{eqnarray}
Performing the Jordan--Wigner transformation one can write the 
Hamiltonian
in terms of fermionic creation and annihilation operators (for more 
details see section \ref{sec:JordWig}):
    \begin{eqnarray}
H_{gl(2|1) \oplus gl(2)} &\!\!=\!\!& \sum_{\atopn{x=1}{\alpha = \uparrow, 
\downarrow}}^L \Big\{
c^{\dagger}_{\alpha \, x+1} c_{\alpha \, x} + c^{\dagger}_{\alpha \, 
x} 
c_{\alpha \, x+1} \Big\} + u \sum_{x=1}^L (1- 2n^{c}_{\downarrow \, x})
(1-2n^{c}_{\uparrow \, x} ) \nonumber \\
&& + \; \sum_{x=1}^L\Big\{(c^{\dagger}_{\uparrow \, x+1 }c_{\uparrow \, 
x} +
c^{\dagger}_{\uparrow \, x} c_{\uparrow \, x+1}) ( n^{d}_{\uparrow \, 
x}
n^{d}_{\uparrow \, x+1} - n^{d}_{\uparrow \, x} - n^{d}_{\uparrow \, 
x+1} )
\nonumber \\
&&\qquad\qquad + \; c^{\dagger}_{\uparrow \, x+1 } \, c_{\uparrow \, 
x} \,
d^{\dagger}_{\uparrow \, x+1 } \, d_{\uparrow \, x} + 
c^{\dagger}_{\uparrow
\, x} \, c_{\uparrow \, x+1} \, d^{\dagger}_{\uparrow \, x } \, 
d_{\uparrow
\, x+1}\Big\} \nonumber \\
&& - \; u \sum_{x=1}^L (1- 2n^{c}_{\downarrow \, x}) (
1- n^{c}_{\uparrow \, x} ) n^{d}_{\uparrow \, x} 
\label{gl21gl2 Ham cao}
    \end{eqnarray}
where $n^{b}_{\alpha \, x} = b^{\dagger}_{\alpha \, x} b_{\alpha \, 
x}$ is the
particle number operator for $b = c,d$, and we assume standard
relations between the operators ($\alpha,\beta = 
\uparrow,\downarrow$):
\begin{equation}
\{ c^{\dagger}_{\alpha \, x}, c_{\beta \, y} \} = \delta_{xy} \;
\delta_{\alpha \beta} \mb{;}
\{ d^{\dagger}_{\alpha \, x}, d_{\beta \, y} \} = \delta_{xy} \;
\delta_{\alpha \beta} \mb{;}
\{ c_{\alpha \, x}, d_{\beta \, y} \} = 0 \,.
\end{equation}

In the next subsection we present the system of Bethe equations
corresponding to this Hamiltonian. Then, in the following 
subsections,
we explain in details the method we used for our calculations:
in subsection \ref{sec:gl21-BAlevel1} we give the description of the 
coordinate Bethe
ansatz approach for the first step of "nested" $gl(2|1) \oplus gl(2)$
problem. Next subsections consist the explanations of the second and 
third steps of the problem.

\subsection{Result for $gl(2|1) \oplus gl(2)$\label{sec:gl21-BAresu}}
As we shall see in the construction of the Bethe ansatz, the model 
describes three different kinds  of "particles" ($e^{2\uparrow}$, 
$e^{2\downarrow}$ and $e^{3\uparrow}$) above a "vacuum". The 
particles $e^{2\uparrow}$ and 
$e^{2\downarrow}$ can be associated to spin up and down 
electrons, while $e^{3\uparrow}$ represents a "spin 0" fermion (see 
section \ref{sec:JordWig} for more details). 

The energy of a state with $N$ excitations is given by
\begin{equation}
E = L-2N + 2 \sum^{N}_{m=1} \cos k_{m} 
\end{equation}
and its momentum reads
\begin{equation}
\fp = \sum^{N}_{m=1} k_{m} 
\end{equation}
where the "impulsions" (or Bethe paramaters) $k_{m}$ obey
the Bethe equations of our $gl(2|1) \oplus gl(2)$ model:
\begin{eqnarray}
&& e^{ik_{j}L}=(-1)^{K+N+1} \prod^{K}_{m=1} \frac{i \sin k_{j}+ia_{m} 
+
\frac{u}{4} }{i \sin k_{j}+ia_{m} - \frac{u}{4}} \;, \quad 
j=1,\ldots,N \\
&& (-1)^{N} \prod_{j=1}^{N} \frac{i \sin k_{j}+ia_{m} + 
\frac{u}{4} }{i
\sin k_{j}+ia_{m} - \frac{u}{4}} = \Lambda(\vec{n}) \prod_{l=1,\; 
l\neq
m}^{K} \frac{ ia_{m} - ia_{l} + \frac{u}{2}}{ia_{m} - ia_{l} - 
\frac{u}{2}}
\;, \quad m=1,\ldots,K\quad \\
&& \Lambda(\vec{n}) = e^{ \frac{2i \pi}{K} (n_{1}+ \ldots 
+n_{M})}\,,
\quad M=0,\ldots,K\,,\quad
 1\leq n_{1}<n_{2}< \ldots <n_{M}\leq K
\end{eqnarray}
where L is the number of sites considered in Hubbard model, N is total
number of $e^{2\uparrow}$, $e^{2\downarrow}$ and $e^{3\uparrow}$
"particles". K counts the total number of excitations $e^{2\uparrow}$
and $e^{3\uparrow}$ and finaly M numbers $e^{3\uparrow}$ "particles".
The integers $n_{j}$ correspond to the Bethe parameter of the last 
level, but their Bethe equation has been solved: it just corresponds 
to 
the quantization of these parameters (see details in section 
\ref{sec:glnm-BAlevel3} that deals with a more general case).

\subsection{Coordinate Bethe ansatz, level one\label{sec:gl21-BAlevel1}}

We use the coordinate Bethe ansatz to find the eigenvalues and 
eigenvectors
of this model. In this model we have 5 types of distinct excitations 
noted
$e^{1\uparrow}, e^{2\uparrow}, e^{3\uparrow}$ and
$e^{1\downarrow}, e^{2\downarrow}$. The objects $e^{a\alpha}_{x}$
are elementary vectors (with $1$ in $a$ position and $0$ elsewhere) 
which
form a natural basis for the vector space $V_{\alpha \, x} \sim \CC^5$
($\alpha$ and $x$ hold to describe spin and position in physical 
space).
First of all we define the reference state (pseudo-vacuum):
\begin{equation}
\phi_{0} = \prod^{L}_{x=1} e^{1\uparrow}_{x} e^{1\downarrow}_{x}\,.
\end{equation}

In the following calculations we use the expression of the 
Hamiltonian only
in matrix form (\ref{gl21gl2 Ham}). The corresponding eigenvalue is 
given by:
\begin{equation}
H_{gl(2|1) \oplus gl(2)}\ \phi_{0} = u L\, \phi_{0}\,.
\end{equation}
We define an excitation by a pair $(A,\alpha)$, $A = 2$ or $3$
(corresponding to vectors $e^{2}$ or $e^{3}$) and $\alpha =
\uparrow,\downarrow$ and there is no pair $(3,\downarrow)$. The $N$
excitation states of the Hamiltonian can be written as
    \begin{equation}
\phi[(\overline{A},\overline{\alpha})] = \sum_{\mathbf{x}}
\Psi[\mathbf{x},(\overline{A},\overline{\alpha})]\,
e^{A_{1}\;\alpha_{1}}_{x_{1}} \ldots e^{A_{N}\; \alpha_{N} }_{x_{N}}
\label{eq:excited_st_gl21gl2}
    \end{equation}
where the sum over $\mathbf{x} = (x_{1},x_{2}, \ldots, x_{N})$ is
considered without coinciding points $x_{l} = x_{m}$ such that
$\alpha_{l}=\alpha_{m}$ for any $l,m = 1, \ldots, N$ (this describes 
the
exclusion principle for identical particles). We also use the notation
$(\overline{A},\overline{\alpha})=(A_{1},\alpha_{1}) \ldots
(A_{N},\alpha_{N})$, and we omitted the `empty sites', i.e. sites 
carrying vectors $e^{1\uparrow} e^{1\downarrow}$.

Applying the Hamiltonian to (\ref{eq:excited_st_gl21gl2}), one gets 
the
eigenvalue equation for the $\Psi$ function, given by
    \begin{eqnarray}
&& \sum^{N}_{m=1} \left( \Psi[\mathbf{x} - \mathbf{e}_{m},
(\overline{A},\overline{\alpha})] \Delta^{-}_{m} + \Psi[\mathbf{x} +
\mathbf{e}_{m},(\overline{A},\overline{\alpha}) ] \Delta^{+}_{m} 
\right) +
\nonumber \\
&& + \Big((L-2N) + u \sum_{l,n} \delta(x_{l}-x_{n}) 
\delta(\alpha_{l}\neq
\alpha_{n}) - E \Big)\Psi[\mathbf{x},(\overline{A},\overline{\alpha})]
\Delta^{3} = 0
\label{psi eq gl21gl2}
    \end{eqnarray}
where $\mathbf{e}_{m}$ is an elementary vector in $\CC^N$ with entry 
$1$ on
the $m^{th}$ position and $0$ elsewhere. Also we denoted
\bea
\Delta^{\pm}_{m} &=& \prod_{l\neq m} \prod_{n \neq m} 
\delta^{\updownarrow}(x_{l} \neq x_{n})\
 \prod_{l} \delta^{\updownarrow}(x_{l} \neq x_{m})
\delta^{\updownarrow}(x_{l} \neq x_{m} \pm 1), \\
\Delta^{3} &=& \prod_{l\neq n} 
\delta^{\updownarrow}(x_{l} \neq x_{n}), 
\\
\delta^{\updownarrow}(x_{l} \neq x_{n}) &=& 1 -
\delta(x_{l}-x_{n})\delta(\alpha_{l}-\alpha_{n})\,.
\label{symbols gl21gl2}
\eea
All these symbols mean that there is no particle with the same spin 
on the
same and neighbouring sites with some conditions corresponding to each
symbol (exclusion principle).

It is convenient in the following to denote
$\Psi[\mathbf{x},(\overline{A},\overline{\alpha})]$ simply by
$\Psi(\mathbf{x})$ when there is no ambiguity.

At the first step, we take a non-interacting regime, which implies $x_{1}
\ll x_{2} \ll \ldots \ll x_{N}$ (in other words, the particles are far
enough from each other). As a consequence, all symbols in (\ref{symbols
gl21gl2}) are equal to $1$ and there is no interaction term. We look for a
solution of this equation in a form of "free particles", namely
$\Psi(\mathbf{x}) \propto e^{i \mathbf{k}\mathbf{x}}$, where $\{k_{1},
k_{2}, \ldots, k_{N}\}$ denote a set of unequal numbers.

We get therefore the value of energy
\begin{equation}
E = L-2N + 2 \sum^{N}_{m=1} \cos k_{m} \,.
\label{free energie gl21gl2}
\end{equation}

In order to consider all other cases of the particle dispositions, we 
assume
the {\itshape Bethe hypothesis} for the general solution of $\Psi(\mathbf{x})$. 
We
divide the coordinate space $(x_{1},x_{2},\ldots,x_{N})$ into $N!$ 
sectors: for
$x_{q_{1}} < x_{q_{2}} < \ldots < x_{q_{N}}$,
\begin{equation}
\Psi_{Q}(\mathbf{x}) = \sum_{P} (-1)^{[sg(Q)]} \Phi(P, Q P^{-1}) e^{i 
P\mathbf{k}\cdot Q\mathbf{x}} 
\label{Psi ansatz gl21gl2}
\end{equation}
where $P=[p_{1},p_{2},\ldots,p_{N}]$ and 
$Q=[q_{1},q_{2},\ldots,q_{N}]$ are
two elements of the permutation group $\fS_N$ and
$P\mathbf{k}\cdot Q\mathbf{x} = \sum_{i} k_{p_{i}} x_{q_{i}}$. 
The symbol $(-1)^{[sg(Q)]}$ stands for the signature of the $Q$-permutation 
when restricted to grade 1 particles (that is $e^{3 \uparrow}$).
For instance, we have the property (valid for any  permutation $Q$ and 
any permutation $\Pi_{ii+1}$):   
$(-1)^{[sg(Q \Pi_{ii+1})]}=(-1)^{[sg(Q)]+[A_{i}][A_{i+1}]}$. We recall that
$\Psi(\mathbf{x})$ and accordingly $\Phi(P,QP^{-1})$ depend on the 
type of
excitations $ (\overline{A},\overline{\alpha})$.

The coefficients $\Phi(P, Q P^{-1})$ are not all independent and
the application of the Hamiltonian represented in equation
(\ref{psi eq gl21gl2}) can reduce their number in several cases.

1. Let us consider the sector $x_{q_{1}} \ll \ldots \ll x_{q_{i}} <
x_{q_{i+1}} \ll x_{q_{N}}$ with $x_{q_{i}} = x_{q_{i+1}} - 1$ and
$\alpha_{q_{i}} = \alpha_{q_{i+1}}$. In this case
$\Delta^{-}_{q_{i+1}}=0$ and $\Delta^{+}_{q_{i}}=0$, all other
symbols in (\ref{symbols gl21gl2}) are equal to $1$. Thus, equation 
(\ref{psi
eq gl21gl2}) becomes
\begin{equation}
\sum_{m \neq q_{i},q_{i+1}} \Big( \Psi_{Q}(\mathbf{x} -
\mathbf{e}_{m})  + \Psi_{Q}(\mathbf{x} + \mathbf{e}_{m}) \Big) +
\Psi_{Q}(\mathbf{x} - \mathbf{e}_{q_{i}}) + \Psi_{Q}(\mathbf{x} +
\mathbf{e}_{q_{i+1}}) = \big(E-L+2N \big)\Psi_{Q}(\mathbf{x})
\end{equation}

Using (\ref{Psi ansatz gl21gl2}) and performing simple algebraic
calculation we obtain
\begin{equation}
\Phi(\Pi_{a b} P, Q P^{-1} \Pi_{ab} ) = - \Phi( P, Q P^{-1} ),
\label{Phi'=-Phi gl21gl2}
\end{equation}
where $\Pi_{ab}$ is the permutation of objects $a$ and $b$ which are 
linked
to $i$,$i+1$ by $p^{-1}(a) = i$ and $p^{-1}(b) = i+1$. This relation 
holds
for any value of $A_{q_{i}}$ and $A_{q_{i+1}}$.

2. Now let us consider another case: $x_{q_{1}} \ll \ldots \ll 
x_{q_{i}} =
x_{q_{i+1}} \ll x_{q_{N}}$ and $\alpha_{q_{i}} \neq 
\alpha_{q_{i+1}}$. We
denote by $\bar{Q}$ the sector where $i$ and $i+1$ are permuted (i.e.
$\bar{Q}=Q \Pi_{i i+1}$). It implies that the equation (\ref{psi eq
gl21gl2}) becomes
\begin{eqnarray}
&& \sum_{m \neq q_{i},q_{i+1}} \left[ \Psi_{Q}(\mathbf{x} - 
\mathbf{e}_{m})
+ \Psi_{Q}(\mathbf{x} + \mathbf{e}_{m}) \right] + \Psi_{Q}(\mathbf{x} 
-
\mathbf{e}_{q_{i}}) + \Psi_{\bar{Q}}(\mathbf{x} + \mathbf{e}_{q_{i}}) 
+ 
\nonumber \\
&& + \;\; \Psi_{\bar{Q}}(\mathbf{x} - \mathbf{e}_{q_{i+1}}) +
\Psi_{Q}(\mathbf{x} + \mathbf{e}_{q_{i+1}}) + \left((L-2N) - E
\right)\Psi_{Q}(\mathbf{x}) = 0\,.
\end{eqnarray}
On the other hand, we have the condition of continuity of the wave function  
\begin{equation}
\Psi_{Q}(\mathbf{x}) = \Psi_{\bar{Q}}(\mathbf{x}) \quad \mbox{with} 
\quad
x_{q_{i}} = x_{q_{i+1}}
\end{equation}
These last two equations together give two conditions on the 
coefficients
$\Phi(P,QP^{-1})$ which can be written in matrix form
\begin{equation}
 \begin{pmatrix}
  \Phi(\Pi_{ab}P, QP^{-1}) \\
  \Phi(\Pi_{ab}P, QP^{-1}\Pi_{ab})
  \end{pmatrix} =
\begin{pmatrix}
  t_{ab} & r_{ab}  \\
  r_{ab} & t_{ab}
  \end{pmatrix}
\begin{pmatrix}
 \Phi(P, QP^{-1}) \\
  \Phi(P, QP^{-1}\Pi_{ab})
  \end{pmatrix}
  \label{Phi=MPhi gl21gl2}
\end{equation}
with
\begin{equation}
t_{ab} = \frac{2i(\lambda_{a} - \lambda_{b} )}{u + 2i (\lambda_{a} -
\lambda_{b}) }, \; \; r_{ab} = \frac{ - u }{u + 2i (\lambda_{a} -
\lambda_{b}) }\,.
  \label{r,t gl21gl2}
\end{equation}
These equations also hold for any type of excitations (any value of
$A_{q_{i}}$ and $A_{q_{i+1}}$). We set $\lambda_{a} = \sin k_{a}$ to
simplify the expressions.

Before applying the periodic conditions on the coefficients
$\Phi(P,QP^{-1})$ (henceforth we will rename $QP^{-1}$ simply by $Q$), we
rewrite equations (\ref{Phi'=-Phi gl21gl2}) and (\ref{Phi=MPhi gl21gl2}) in a
compact form. Let us introduce the vector
\begin{equation}
\hat{\Phi}(P) \equiv \sum_{Q,(\overline{A},\overline{\alpha})}
\Phi[P,Q,(\overline{A},\overline{\alpha})]
|(A_{q_{1}},\alpha_{q_{1}}),\ldots,(A_{q_{N}},\alpha_{q_{N}})>
\end{equation}
where the sum is over all types of excitations and all 
corresponding
sectors. The vector
$|(A_{q_{1}},\alpha_{q_{1}}),\ldots,(A_{q_{N}},\alpha_{q_{N}})> $ 
belongs
to $V_{1} \otimes \ldots \otimes V_{N}$, where $V=\spn\{
2\uparrow,2\downarrow,3\uparrow\}$, and represents one state in the 
space of $N$
excitations. Let us illustrate by an example for $N=2$ excitations: 
in this
case we have $6$ types of different excitations.
\begin{eqnarray}
&& \sum_{Q,(\overline{A},\overline{\alpha})}
\Phi[P,Q,(\overline{A},\overline{\alpha})]
|(A_{q_{1}},\alpha_{q_{1}}),(A_{q_{2}},\alpha_{q_{2}})> =
\Phi(P,id,1)|2\uparrow,2\uparrow> + 
\Phi(P,id,2)|2\uparrow,2\downarrow>
\nonumber \\
&& \qquad + \Phi(P,id,3)|2\uparrow,3\uparrow> +
\Phi(P,\Pi_{12},2)|2\downarrow,2\uparrow> + 
\Phi(P,id,4)|2\downarrow,2\downarrow> + \nonumber \\
&& \qquad + \Phi(P,id,5)|2\downarrow,3\uparrow> +
\Phi(P,\Pi_{12},3)|3\uparrow,2\uparrow> + 
\Phi(P,\Pi_{12},5)|3\uparrow,2\downarrow> \nonumber \\
&& \qquad + \Phi(P,id,6)|3\uparrow,3\uparrow>\,.
\end{eqnarray}
Thus, for $N=2$ excitations we can express (\ref{Phi'=-Phi gl21gl2}) 
and
(\ref{Phi=MPhi gl21gl2}) by
\begin{eqnarray}
\hat{\Phi}(\Pi_{12}P) &=& (-1)\Phi(P,id,1)|2\uparrow,2\uparrow> +
\left[t_{12}\Phi(P,id,2) + r_{12} \Phi(P,\Pi_{12},2)
\right]|2\uparrow,2\downarrow> \nonumber \\ 
&& + \left[t_{12}\Phi(P,\Pi_{12},2) + r_{12} \Phi(P,id,2) \right]
|2\downarrow,2\uparrow> + (-1) \Phi(P,id,3)|2\uparrow,3\uparrow> 
\nonumber \\ 
&& +  (-1)\Phi(P,\Pi_{12},3)|3\uparrow,2\uparrow> + \text{\ldots} +
(-1)\Phi(P,id,6)|3\uparrow,3\uparrow> \nonumber \\
&\equiv& S^{(1)}_{12}(\lambda_{1}-\lambda_{2}) \hat{\Phi}(P)
\end{eqnarray}
where we introduced the so-called $S$-matrix $S^{(1)}_{12}(\lambda)$:
\begin{equation}
S^{(1)}_{12}(\lambda_{1}-\lambda_{2})
\begin{pmatrix}
 |2\uparrow, 2\uparrow> \\
 |2\uparrow, 2\downarrow> \\
 |2\uparrow, 3\uparrow> \\
 \hline
 |2\downarrow, 2\uparrow> \\
 |2\downarrow, 2\downarrow> \\
 |2\downarrow, 3\uparrow> \\
 \hline
 |3\uparrow, 2\uparrow> \\
 |3\uparrow, 2\downarrow> \\
 |3\uparrow, 3\uparrow> \\
  \end{pmatrix}
 =
\left(
\begin{array}{ccc|ccc|ccc}
  -1 & & & & & & & & \\
  & t_{12}& & r_{12} & & & & & \\
  & & 0 & & & & -1 & & \\
 \hline
  & r_{12}& & t_{12} & & & & & \\
  & & & & -1 & & & & \\
  & & & & & t_{12} & & r_{12} & \\
 \hline
  & & -1 & & & & 0 & & \\
  & & & & & r_{12} & & t_{12} & \\
  & & & & & & & & -1 \\
  \end{array}
\right)
\begin{pmatrix}
 |2\uparrow, 2\uparrow> \\
 |2\uparrow, 2\downarrow> \\
 |2\uparrow, 3\uparrow> \\
 \hline
 |2\downarrow, 2\uparrow> \\
 |2\downarrow, 2\downarrow> \\
 |2\downarrow, 3\uparrow> \\
 \hline
 |3\uparrow, 2\uparrow> \\
 |3\uparrow, 2\downarrow> \\
 |3\uparrow, 3\uparrow> \\
  \end{pmatrix}
  \label{S12 gl21gl2}
\end{equation}
The coefficients $t_{12}$ and $r_{12}$ are defined in (\ref{r,t gl21gl2}).
The same reasoning could be repeated for an arbitrary number $N$
of excitations, and we obtain
\begin{equation}
\hat{\Phi}(\Pi_{ab}P) = S^{(1)}_{ab}(\lambda_{a}-\lambda_{b})
\hat{\Phi}(P),
\end{equation}
where the matrix $S^{(1)}_{ab}(\lambda_{a}-\lambda_{b})$ acts 
nontrivialy
only on vector spaces $V_{a} \otimes V_{b}$. It satisfies the 
Yang--Baxter
equation:
\begin{equation}
S^{(1)}_{12}(\lambda_{1}-\lambda_{2})
S^{(1)}_{13}(\lambda_{1}-\lambda_{3})
S^{(1)}_{23}(\lambda_{2}-\lambda_{3})
=S^{(1)}_{23}(\lambda_{2}-\lambda_{3})
S^{(1)}_{13}(\lambda_{1}-\lambda_{3})
S^{(1)}_{12}(\lambda_{1}-\lambda_{2})\,.
\end{equation}

3. In order to obtain the Bethe equations we should apply the periodic
boundary conditions on the function $\Psi(\mathbf{x})$. Let $C$ be the
cyclic permutation given by $C=\Pi_{N 1} \ldots \Pi_{N N-1}$. More
precisely the periodicity condition means
\begin{equation}
\Psi_{QC}(\mathbf{x}+\mathbf{e}_{q_{1}}L)=\Psi_{Q}(\mathbf{x})
\end{equation}
which implies a condition on the coefficients $\Phi(P,Q)$, namely
\begin{equation}
\hat{\Phi}(PC) = e^{ik_{p_{N}}L}\hat{\Phi}(P) \label{periodicity 
gl21gl2}
\end{equation}
choosing $P=C^{N-j}$, one derives a system of equations satisfied by 
the
coefficients $\hat{\Phi}(id)$ which is called "auxiliary problem".
For $j=1,\ldots,N$, it reads
\begin{equation}
S^{(1)}_{j+1 j} \ldots S^{(1)}_{N j} S^{(1)}_{1 j} \ldots 
S^{(1)}_{j-1 j}
\hat{\Phi}(id) = e^{ik_{j}L} \hat{\Phi}(id) \label{AuxPr1 gl21gl2}
\end{equation}
here we omitted the arguments in $S$-matrices, $ S^{(1)}_{a b} \equiv
S^{(1)}_{a b}(\lambda_{a}-\lambda_{b})$.

If we could perform the diagonalization of the left-hand-side in the
general case, it would be possible right now to write the Bethe equations
for our model. The form of matrix $S^{(1)}_{ab}$ requires to use again the
coordinate Bethe ansatz to solve the auxiliary problem.

\subsection{Auxiliary problem, level two\label{sec:gl21-BAlevel2}}

Again we have the equation on eigenvectors and eigenvalues to
solve, but contrarily to the first step the Hamiltonian is a little
more complicated:
\begin{equation}
S^{(1)}_{j+1 j} \ldots S^{(1)}_{N j} S^{(1)}_{1 j} \ldots 
S^{(1)}_{j-1 j}
\phi = \Lambda_{j} \phi \label{SSSphi=Lphi gl21gl2}
\end{equation}
Since the matrix $S^{(1)}$ is regular, the above Hamiltonian can be 
identified with $\wt t(\lambda_{j})$, where we have introduced a new 
transfer matrix
$$
\wt t(\lambda)=tr_{0}\Big(
S^{(1)}_{j+1,0}(\lambda_{j+1}-\lambda)\,
\ldots 
S^{(1)}_{N,0}(\lambda_{N}-\lambda)\,S^{(1)}_{1,0}(\lambda_{1}-\lambda)\,
\ldots 
S^{(1)}_{j-1,0}(\lambda_{j-1}-\lambda)\,S^{(1)}_{j,0}(\lambda_{j}-\lambda)
\Big)
$$
Since $S^{(1)}$ obeys Yang--Baxter equation, this new problem is indeed
integrable. We can thus perform the Bethe ansatz again.

First, we slightly modify the $S$-matrix $S^{(1)}_{12}\rightarrow -
S^{(1)}_{12} \equiv S_{12}$ to simplify the following
calculations. Thus, it is given by
\begin{eqnarray}
S_{12}(\lambda_{1} - \lambda_{2}) &=& P_{12} - t_{12}K_{12}\,, 
\mb{with} t_{12}=\frac{i(\lambda_{1} - \lambda_{2} )}{i
(\lambda_{1} - \lambda_{2}) + \frac{u}{2}}\,,
\\
 P_{12} &=& \sum_{i,j=1,2,3}E^{ij} \otimes E^{ji}\,, \\
K_{12} &=& \sum_{i=2,3} \left( E^{11} \otimes E^{ii}+ E^{ii}
\otimes E^{11} + E^{1i} \otimes E^{i1} + E^{i1} \otimes E^{1i} 
\right)\,.
\end{eqnarray}
We  use the coordinate Bethe ansatz to proceed with this problem. 
At
this step we have only three types of different excitations: 
$e^{2\uparrow},
e^{2\downarrow}$ and $e^{3\uparrow}$. At first, we choose the
reference state (pseudovacuum) to be
$$
\phi_{0} = \prod^{N}_{k=1} e^{2\downarrow}_{k},
$$
with eigenvalue $\Lambda = 1$. \\
Applying the $S$-matrix on different states (on the pseudovacuum 
$e^{2}$ and
two excitations $e^{1}$ and $e^{3}$), we get
\begin{eqnarray}
&& S_{12}(\lambda_{1} - \lambda_{2})\,e^{2\downarrow} \otimes 
e^{2\downarrow} 
= e^{2\downarrow} \otimes e^{2\downarrow}, \\
&& S_{12}(\lambda_{1} - \lambda_{2})\,e^{A} \otimes e^{2\downarrow} = 
-t_{12}\,
e^{A}\otimes e^{2\downarrow} + (1-t_{12})\,e^{2\downarrow} \otimes 
e^{A}, \\
&& S_{12}(\lambda_{1} - \lambda_{2})\,e^{2\downarrow} \otimes e^{A} = 
-t_{12}\,
e^{2\downarrow}\otimes e^{A} + (1-t_{12})\,e^{A} \otimes 
e^{2\downarrow}, \\
&& S_{12}(\lambda_{1} - \lambda_{2})\,e^{A} \otimes e^{B} = e^{B} 
\otimes
e^{A}\mb{with}A,B=2\uparrow,\,3\uparrow\,. 
\end{eqnarray}
We first consider the "non-interacting" regime, which implies to
solve the equation with only one excitation. Let $A = 2\uparrow$ or 
$3\uparrow$
and indicate the type of the
excitation
\begin{equation}
\phi(A)=\sum_{x} f_{x,A} \; e^{A}_{x}\,.
\label{def: f func}
\end{equation}

We will forget henceforth the $A$ index to simplify the notations.

Here, contrarily to the first subsection, it
is not obvious to readily calculate the Hamiltonian action on
excited states. We can do it step by step introducing a recursive
Hamiltonian (for details, see \cite{sutherland})
\begin{equation}
S_{j-k j} \ldots S_{j-1 j} \phi = \Lambda_{j} \phi \label{recursHam 
gl21gl2}
\end{equation}
and recursive coefficients $f^{(k)}_{x}$ which represent the action of the
recursive Hamiltonian on the coefficients $f_{x}$. One gets the following
relations
\begin{eqnarray}
&& \sum_{x} f^{(1)}_{x} e^{A}_{x} \equiv S_{j-1 j} \sum_{x} f_{x}
e^{A}_{x}, \nonumber \\
&& \vdots \\
&& \sum_{x} f^{(k)}_{x} e^{A}_{x} \equiv S_{j-k j} \ldots S_{j-1 j}
\sum_{x} f_{x} e^{A}_{x}\,. \nonumber
\end{eqnarray}
We write down the recursive relations between coefficients
$f^{(k)}_{x}$ and $f^{(k-1)}_{x}$. For $x\neq j,j-k$
\begin{eqnarray}
&& f^{(k)}_{x} = f^{(k-1)}_{x}\,, \nonumber \\
&& f^{(k)}_{j} = -t_{j-k j} f^{(k-1)}_{j} + (1-t_{j-k 
j})f^{(k-1)}_{j-k}\,, \\
&& f^{(k)}_{j-k} = -t_{j-k j} f^{(k-1)}_{j-k} + (1 -t_{j-k 
j})f^{(k-1)}_{j} \,.
\nonumber
\end{eqnarray}
Skipping the details of calculation, the solution of the equation
(\ref{recursHam gl21gl2}) gives a relation between different $f_{x}$. 
Closer considering shows that one needs to introduce an additional 
constant $a$
(particle rapidity), such that
\begin{equation}
\frac{f_{x+1}(a)}{f_{x}(a)} = - \frac{i \lambda_{x} + i a +
\frac{u}{4}}{i \lambda_{x+1} + i a - \frac{u}{4}}\ .
\end{equation}
Thus, we can write down the expression for coefficient $f_{x}$ and the
eigenvalue corresponding to one excitation. We normalize $f_{1} = 1$, 
then we
have
\begin{equation}
f_{x}(a) = \prod^{x-1}_{m=1} \left(- \frac{i \lambda_{m} + i a +
\frac{u}{4}}{i \lambda_{m+1} + i a - \frac{u}{4}}\right),
\label{f gl21gl2}
\end{equation}
\begin{equation}
\Lambda_{j} \equiv \sigma_{j}(a) = - \frac{i \lambda_{j} + i a +
\frac{u}{4}}{i \lambda_{j} + i a - \frac{u}{4}}\,. 
\label{sigma gl21gl2}
\end{equation}
It is convenient to write down some relation between iterated
coefficient $f^{(k)}_{x}$, $f_{x}$ and $\sigma$:
\begin{eqnarray}
&& f_{x}(a) = \sigma_{1}(a) \ldots \sigma_{x-1}(a) \; \frac{i 
\lambda_{1} + i
a - \frac{u}{4}}{i \lambda_{x} + i a - \frac{u}{4}}\,, \nonumber \\
&& f^{(k)}_{j}(a) = \frac{f_{j}(a)}{\sigma_{j-1}(a) \ldots 
\sigma_{j-k}(a)}\,, 
\label{fk,f,sigma gl21gl2}\\
&& f^{(k)}_{j-k}(a) = \sigma_{j}(a)f_{j-k}(a)\,. \nonumber
\end{eqnarray}
Starting from the initial equation (\ref{SSSphi=Lphi gl21gl2}), we can
choose $j=N$ and without any loss of generality we can derive the equation
which represents the Bethe equations on rapidity $a$ by application of the
initial Hamiltonian on $f_{x}(a)$. It also automatically implies the
periodic boundary condition
\begin{equation}
  \prod^{N}_{m=1} \sigma_{m}(a) = 1\,.
\end{equation}
All these results hold for any type of excitation $e^{2\uparrow}$ or
$e^{3\uparrow}$ (we neglected the index A above).

Now we can proceed to the case of K excitations. Let $\bar{A}$ be
a vector $(A_{1}, \ldots ,A_{K})$ with $A_{i}=2\uparrow,3\uparrow$, 
which represents the
combination of different excitations in consideration. The
eigenvector is naturally represented as
\begin{equation}
\phi(\bar{A}) = \sum_{\mathbf{x}} \Psi(\mathbf{x}, \bar{A})
e^{A_{1}}_{x_{1}} \ldots e^{A_{K}}_{x_{K}}\,.
\end{equation}
The sum is done over all coordinates $x_{i}$ without coinciding
points $x_{i}=x_{k}$ for any $i,k$ (exclusion principle).

We use the Bethe hypothesis for the coefficients
$\Psi(\mathbf{x},\bar{A})$, in the sector $x_{q_{1}}<x_{q_{2}}< \ldots
<x_{q_{K}}$ where $Q=[q_{1}, \ldots ,q_{K}]$ is a permutation of the
integers $1,2, \ldots ,K$:
\begin{equation}
\Psi(\mathbf{x},\bar{A}) = \sum_{P} \Phi(P,QP^{-1}, \bar{A})
f_{x_{q_{1}}}(a_{p_{1}}) f_{x_{q_{2}}}(a_{p_{2}})
 \ldots f_{x_{q_{K}}}(a_{p_{K}})
\end{equation}
where $f_{x_{q_{i}}}(a_{p_{i}})$ is the one-particle solution with rapidity
$a_{p_{i}}$. We will omit the index $\bar{A}$ to simplify the notations. As
above, one can reduce the number of coefficients $\Phi(P,QP^{-1})$.

At this stage we will consider K excitations of different types. Therefore
we have $K!/M!(K-M)!$ sectors $x_{q_{1}}<x_{q_{2}}< \ldots <x_{q_{K}}$,
where $M$, $K-M$ are the numbers of excitation of type $2\uparrow$ and
$3\uparrow$ respectively. Acting with the recursive Hamiltonian on the
eigenvector $\phi$, see (\ref{recursHam gl21gl2}), we are able to write down
the relations between different $\Psi^{(k)}_{Q}(\mathbf{x})$.

 We consider
$j$ to take the biggest value such that we do not have any excitation on
the sites bigger than $j$. Several cases can occur, depending on 
whether there is an excitation on sites $j$ and/or $j-k$. If there is 
an excitation on site $j$, it must correspond to $x_{q_{K}}=j$. For 
the site $j-k$, there exists $q_{m}$ such that we have either
$x_{q_{m}}<j-k<x_{q_{m+1}}$ (no excitation at site $j-k$), or 
$x_{q_{m}}=j-k$ (one excitation at site $j-k$). Then, we have
\begin{equation}
\Psi^{(k)}_{Q}(\mathbf{x}) = \Psi^{(k-1)}_{Q}(\mathbf{x}), \ 
\text{for}\quad \mathbf{x}\neq j,j-k,
\end{equation}
\beano
&&\Psi^{(k)}_{Q}(  \ldots ,
\mbox{\raisebox{1.2ex}{${\substack{x_{q_{K}}\\ \downarrow \\ \textstyle
j}}$}}, 
\ldots  ) 
= -t_{j-k j}
\Psi^{(k-1)}_{Q}( \ldots ,
\mbox{\raisebox{1.2ex}{${\substack{x_{q_{K}}\\ \downarrow \\ \textstyle
j}}$}}
, \ldots ) + (1-t_{j-k j})
\Psi^{(k-1)}_{Q'}( \ldots ,
\mbox{\raisebox{1.2ex}{${\substack{x_{q_{K}}\\ \downarrow \\ \textstyle
j-k}}$}}
, \ldots ),
\nonu
&&\mb{with} Q'=Q \Pi_{K m+1} \ldots \Pi_{K K-1} \mb{for} \mathbf{x}\neq j-k,
\\[1.2ex]
&&\Psi^{(k)}_{Q}(\ldots,
\mbox{\raisebox{1.2ex}{${\substack{x_{q_{m}}\\ \downarrow \\ \textstyle
j-k}}$}}
,\ldots) = -t_{j-k j}
\Psi^{(k-1)}_{Q}(\ldots,
\mbox{\raisebox{1.2ex}{${\substack{x_{q_{m}}\\ \downarrow \\ \textstyle
j-k}}$}}
,\ldots) + (1-t_{j-k j})
\Psi^{(k-1)}_{Q'}(\ldots,
\mbox{\raisebox{1.2ex}{${\substack{x_{q_{m}}\\ \downarrow \\ \textstyle
j}}$}}
,\ldots)
\nonu
&&\mb{with} Q'=Q \Pi_{m K} \ldots \Pi_{m m+1}
\mb{for} \mathbf{x}\neq j
\eeano
In this last case, we have also the exclusion principle: for any $n
\neq m$, $x_{n} \neq j,j-k$.

When two excitations are on sites $j$
and $j-k$, with $x_{q_{K}}=j$ and $x_{q_{m}}=j-k$,
 we have:
\beano
\Psi^{(k)}_{Q}\left( \ldots ,j-k, \ldots ,j,\ldots\right) =
\Psi^{(k-1)}_{Q \Pi_{Km}}( \ldots ,j, \ldots ,j-k,\ldots)
\mb{for non-identical excitations,}
\\
\Psi^{(k)}_{Q}\left( \ldots ,j-k, \ldots ,j,\ldots\right) =
\Psi^{(k-1)}_{Q}( \ldots ,j-k, \ldots ,j,\ldots)
\mb{if the excitations are identical.}
\eeano
On the one hand, we have the Bethe hypothesis for the coefficients
$\Psi(\mathbf{x})$ composed of the one-excitation functions $f_{x_{i}}$; on
the other hand, we have the relations for the iterated coefficients, it is
natural to calculate how the iteration passes to the one-excitation
functions. In other words, we want to calculate the coefficients
$\Psi^{(k)}_{Q}(\mathbf{x})$ in terms of the iterated free excitation
functions $f^{(k)}_{x}(a_{p})$.

After first iteration for any $\mathbf{x}$ except the coefficient
$\Psi^{(1)}_{Q}( \ldots ,j-1, \ldots ,j, \ldots )$ we find
\begin{equation}
\Psi^{(1)}_{Q}(\mathbf{x}) = \sum_{P} \Phi(P,QP^{-1})
f^{(1)}_{x_{q_{1}}}(a_{p_{1}}) f^{(1)}_{x_{q_{2}}}(a_{p_{2}})
 \ldots f^{(1)}_{x_{q_{K}}}(a_{p_{K}})\,.
\end{equation}
The next iteration will allows us to make an assumption on the
form of $\Psi^{(1)}_{Q}( \ldots ,j-1, \ldots ,j, \ldots )$. Thus we 
calculate the
coefficients $\Psi^{(2)}_{Q}(\mathbf{x})$ except again those where
two coordinates coincide with $j-2$ and $j$. Skiping the
calculations we find for the first iteration
$$
\Psi^{(1)}_{Q}( \ldots ,j-1, \ldots ,j, \ldots ) = \sum_{P} \Phi(P,(Q 
\Pi_{K K-1}
)P^{-1}) f^{(1)}_{x_{q_{1}}}(a_{p_{1}}) \ldots 
f^{(1)}_{j}(a_{p_{K-1}})
f^{(1)}_{j-1}(a_{p_{K}}).
$$
This implies the conditions
\bea
\Psi^{(1)}_{Q}( \ldots ,j-1,j, \ldots ) &=& \Psi_{Q}( \ldots ,j-1,j, 
\ldots ) \mb{for identical excitations}\\
\Psi^{(1)}_{Q}( \ldots ,j-1,j, \ldots ) &=&
\Psi_{Q'}( \ldots ,j,j-1, \ldots ) \mb{for different
type of excitations}
\eea
with $Q'=Q \Pi_{K K-1}$. It
is useful to notice that for identical excitations $Q \Pi_{K K-1}$
defines the same sector as $Q$. These equations give the connection
between $\Phi(P\Pi_{K K-1},Q(P\Pi_{K K-1})^{-1})$ and
$\Phi(P,QP^{-1})$ for some sectors $Q$. We will consider it
precisely in the general case.

For the second iteration we find
\begin{equation}
\Psi^{(2)}_{Q}(\mathbf{x}) = \sum_{P} \Phi(P,QP^{-1})
f^{(2)}_{x_{q_{1}}}(a_{p_{1}}) \ldots f^{(2)}_{x_{q_{K}}}(a_{p_{K}})
\end{equation}
except
\begin{equation}
\Psi^{(2)}_{Q}( \ldots ,j-1, \ldots ,j, \ldots ) = \sum_{P} 
\Phi(P,(Q\Pi_{K
K-1})P^{-1})
f^{(2)}_{x_{q_{1}}}(a_{p_{1}}) \ldots f^{(2)}_{j}(a_{p_{K-1}})
f^{(2)}_{j-1}(a_{p_{K}}).
\end{equation}
and as in the previous case we have two "undefined" coefficients
$\Psi^{(2)}( \ldots ,j-2, \ldots ,j, \ldots )$ and
$\Psi^{(2)}( \ldots ,j-2, \ldots ,j-1, \ldots ,j, \ldots )$.

Generalizing these results until the $(k+1)$-th iteration, we make 
again a
set of assumptions for the coefficients of type $\Psi^{(k)}( \ldots 
,j-k,
\ldots j, \ldots )$. Hence, for the $k$-th iteration, we find the 
coefficients
\begin{equation}
\Psi^{(k)}_{Q}(\mathbf{x}) = \sum_{P} \Phi(P,QP^{-1})
f^{(k)}_{x_{q_{1}}}(a_{p_{1}}) \ldots f^{(k)}_{x_{q_{K}}}(a_{p_{K}})
\end{equation}
except a set of coefficients. Let $n_{i}$ be integers such that
$0<n_{1}<n_{2}< \ldots <n_{l}<k$, 
$l=1,\ldots,k-1$, that label the position of the possible excitations between 
$j-k$ and $j$. We have a set of 
assumptions made from $(k+1)$-th iteration\footnote{In order to make 
the notation shorter we do not write the dots $\ldots$
for unchanged indices when it is not ambiguous.}. 

If there is an 
excitation in position $j-k$, we have:
\begin{eqnarray}
&& \Psi^{(k)}_{Q}(j-k,j) = \sum_{P} \Phi(P,(Q \Pi^{K-1}_{K})P^{-1})
f^{(k)}_{x_{q_{1}}}(a_{p_{1}}) \ldots f^{(k)}_{j}(a_{p_{K-1}})
f^{(k)}_{j-k}(a_{p_{K}}) \\
&& \Psi^{(k)}_{Q}(j-k,j-n_{1},j) =  \\
&& \quad \sum_{P} \Phi(P,(Q \Pi^{K-2}_{K}
\Pi^{K-1}_{K} ) P^{-1}) f^{(k)}_{x_{q_{1}}}(a_{p_{1}}) \ldots 
f^{(k)}_{j}(a_{p_{K-2}})
f^{(k)}_{j-k}(a_{p_{K-1}}) f^{(k)}_{j-n_{1}}(a_{p_{K}}) \nonu
&& \vdots \nonumber \\
&& \Psi^{(k)}_{Q}(j-k,j-n_{l}, \ldots ,j-n_{1},j) = \\
&& \quad \sum_{P} \Phi(P,(Q\Pi^{K-l-1}_{K} \ldots 
\Pi^{K-1}_{K})P^{-1})
f^{(k)}_{x_{q_{1}}}(a_{p_{1}}) \ldots f^{(k)}_{j}(a_{p_{K-l-1}}) f^{(k)}_{j-k}(a_{p_{K-l}}) 
\ldots
f^{(k)}_{j-n_{1}}(a_{p_{K}}) \nonu
&& \vdots \nonumber \\
&& \Psi^{(k)}_{Q}(j-k,j-k+1 \ldots ,j-1,j) =  \\
&& \quad \sum_{P}
\Phi(P,(Q\Pi^{K-k}_{K} \ldots \Pi^{K-1}_{K})P^{-1})
f^{(k)}_{x_{q_{1}}}(a_{p_{1}}) \ldots f^{(k)}_{j}(a_{p_{K-k}})
f^{(k)}_{j-k}(a_{p_{K-k+1}}) \ldots f^{(k)}_{j-1}(a_{p_{K}})
\nonumber
\end{eqnarray}
If there is no $x$ equal to $j-k$, from the calculation we have
\begin{eqnarray}
&& \Psi^{(k)}_{Q}(j-n_{1},j) = \sum_{P}
\Phi(P,(Q\Pi^{K-1}_{K})P^{-1})
f^{(k)}_{x_{q_{1}}}(a_{p_{1}}) \ldots f^{(k)}_{j}(a_{p_{K-1}})
f^{(k)}_{j-n_{1}}(a_{p_{K}}), \\
&& \Psi^{(k)}_{Q}(j-n_{2},j-n_{1},j) = \nonumber \\
&& \quad \sum_{P}
\Phi(P,(Q\Pi^{K-2}_{K}\Pi^{K-1}_{K})P^{-1})
f^{(k)}_{x_{q_{1}}}(a_{p_{1}}) \ldots f^{(k)}_{j}(a_{p_{K-2}})
f^{(k)}_{j-n_{2}}(a_{p_{K-1}}) f^{(k)}_{j-n_{1}}(a_{p_{K}}), \\
&& \vdots \nonumber \\
&& \Psi^{(k)}_{Q}(j-k+1, \ldots ,j-1,j) =  \\
&& \ \sum_{P}
\Phi(P,(Q\Pi^{K-k+1}_{K} \ldots \Pi^{K-1}_{K})P^{-1})
f^{(k)}_{x_{q_{1}}}(a_{p_{1}}) \ldots f^{(k)}_{j}(a_{p_{K-k+1}})
f^{(k)}_{j-k+1}(a_{p_{K-k+2}}) \ldots  f^{(k)}_{j-1}(a_{p_{K}}) 
\nonumber
\end{eqnarray}
We can write a set of conditions from all made assumptions to
determine the relations between different $\Phi(P,P^{-1})$. One
can see that we can consider only the $k$-th iteration to derive
the necessary relations. More exactly we have $k$ equations to
satisfy
\begin{equation}
\Psi^{(k)}_{Q}(j-k,j-n_{l}, \ldots ,j-n_{1},j) =
\Psi^{(k-1)}_{Q}(j-k,j-n_{l}, \ldots ,j-n_{1},j)
\end{equation}
for identical excitations corresponding to $j$ and $j-k$. 
Remark that in this case $Q \Pi^{K-l-1}_{K}$ is identical to $Q$. 

For non-identical excitations, we have
\begin{equation}
\Psi^{(k)}_{Q}(j-k,j-n_{l}, \ldots ,j-n_{1},j) = \Psi^{(k-1)}_{Q
\Pi^{K-l-1}_{K}}(j,j-n_{l}, \ldots ,j-n_{1},j-k)
\end{equation}
with $l=0,\ldots,k-1$ in both cases. Hence we
have a set of conditions:
$$
\begin{array}{l}
\sum_{P} \Phi(P,(Q\Pi^{K-l-1}_{K} \ldots \Pi^{K-1}_{K})P^{-1})
f^{(k)} \ldots f^{(k)}_{j}(a_{p_{K-l-1}})
f^{(k)}_{j-k}(a_{p_{K-l}}) \ldots f^{(k)}_{x_{j-n_{1}}}(a_{p_{K}}) = 
\\
\quad  = \sum_{P}
\Phi(P,(Q\Pi^{K-l-1}_{K} \ldots \Pi^{K-1}_{K})P^{-1})
f^{(k-1)} \ldots f^{(k-1)}_{j-k}(a_{p_{K-l-1}})
f^{(k-1)}_{j}(a_{p_{K-l}}) \ldots f^{(k-1)}_{x_{j-n_{1}}}(a_{p_{K}}) \,.
\end{array}
$$

Using the expressions obtained for iterated one-particle functions
(\ref{fk,f,sigma gl21gl2}) and omitting all the algebraic calculations, we
have the expression valid for any sector $Q$ and any type of excitations:
\begin{equation}
\frac{\Phi(P,QP^{-1})}{\Phi(P \Pi_{i i+1}, Q(P \Pi_{i i+1})^{-1}
)} = \alpha_{p_{i} p_{i+1}} \equiv  \frac{ia_{p_{i}} -
ia_{p_{i+1}} + \frac{u}{2}}{ia_{p_{i}} - ia_{p_{i+1}} -
\frac{u}{2}} 
\mb{for}i=K-k,\ldots,K-1\,.
\label{phi'=alpha phi gl21gl2}
\end{equation}

Before applying the whole Hamiltonian (\ref{SSSphi=Lphi gl21gl2}), which
also represents the periodic boundary conditions, on the coefficients
$\Phi(P,QP^{-1})$, we rewrite the obtained equation (\ref{phi'=alpha phi
gl21gl2}) in a form gathering all possibilities of excitations
$\overline{A}=(A_{1}, \ldots ,A_{K})$. At first we introduce an object
(henceforth we will rename $QP^{-1}$ simply by $Q$)
\begin{equation}
\hat{\Phi}(P) \equiv \sum_{Q,\overline{A}} \Phi(P,Q,\overline{A})
|A_{q_{1}}, \ldots ,A_{q_{K}}>
\label{def: hatPhi gl21gl2}
\end{equation}
where the sum is over all types of excitations and all
corresponding sectors. The vector $|A_{q_{1}} , \ldots ,A_{q_{K}}> $ 
belongs
to $V_{1} \otimes  \ldots  \otimes V_{K}$ (where 
$V=\spn\{e^{2\uparrow},e^{3\uparrow}\}$)
and represents one combination of $K$ excitations. To explain the
notation we write an example for $K=2$ excitations. In this case
we have $4$ types of different states:
$$
\begin{array}{l}
\sum_{Q,\overline{A}} \Phi(P,Q,\overline{A}) |A_{q_{1}}
,A_{q_{2}}> = \Phi(P,id,1)|2\uparrow,2\uparrow> + 
\Phi(P,id,2)|2\uparrow,3\uparrow> + \\
\qquad \qquad \qquad \qquad \qquad \qquad \quad
\Phi(P,\Pi_{12},2)|3\uparrow,2\uparrow> + 
\Phi(P,id,3)|3\uparrow,3\uparrow>\,.
\end{array}
$$
Thus, for $K=2$ excitations we can express (\ref{phi'=alpha phi 
gl21gl2})
by
$$
\begin{array}{l}
\hat{\Phi}(\Pi_{12}P) = \alpha^{-1}_{1 2} \Big( 
\Phi(P,id,1)|2\uparrow,2\uparrow> 
+ \Phi(P,id,2)|3\uparrow,2\uparrow> + \\
\qquad \qquad \qquad \; \Phi(P,\Pi_{12},2)|2\uparrow,3\uparrow> +
\Phi(P,id,3)|3\uparrow,3\uparrow> \Big) \equiv S^{(2)}_{12} 
\hat{\Phi}(P)
\end{array}
$$
where we introduced, as in the first section, a new
"$S$-matrix", noted $S^{(2)}_{12}$. One can see that this $S$-matrix in
this case is a pure permutation.

The same reasoning could be repeated for an arbitrary number $K$
of excitations and we obtain
\begin{equation}
\hat{\Phi}(\Pi_{ab}P) = \alpha^{-1}_{ab} P_{ab} \hat{\Phi}(P),
\label{S(2)12 gl21gl2}
\end{equation}
where $a=p(i)$, $b=p(i+1)$ and the permutation $P_{ab}$ acts nontrivialy
only on $V_{a} \otimes V_{b}$ vector spaces. Obviously, the matrix
$S^{(2)}_{12}$ being a permutation, satisfies the Yang--Baxter equation.

To obtain the Bethe equations we apply the periodic boundary
conditions which are automatically implied if we pass to the
initial equation (\ref{SSSphi=Lphi gl21gl2}). We choose $j=N$, to
get
$$
\Psi^{(N-1)}_{Q}(\mathbf{x}) = \Lambda_{N}\ \Psi_{Q}(\mathbf{x})\,.
$$
If no $x$ equals $N$, we find the eigenvalue 
$\Lambda_{N} = \sigma_{N}(a_1) \ldots \sigma_{N}(a_K)$
where $\sigma$ is the eigenvalue of one excitation function
(\ref{sigma gl21gl2}).
Otherwise, the equation
$$
\Psi^{(N-1)}_{Q}(x_{1}, \ldots, N, \ldots,x_{K} ) = 
\Lambda_{N} \Psi_{Q}( x_{1}, \ldots, N, \ldots,x_{K} )
$$
leads to additional condition on the coefficients $\Phi(P,Q)$.
Hence, in terms of $\hat{\Phi}(P)$, we have
$$
\hat{\Phi}(PC)= \prod^{N}_{k=1} \sigma_{k}(a_{p_{1}})\ \hat{\Phi}(P)
$$
where $C$ is a cyclic permutation given by $C=\Pi_{K 1} \ldots 
\Pi_{K K-1}$.

Choosing $P=C^{K-m}$, we can derive a system of equations on
coefficients $\hat{\Phi}(id)$ which will be called
"auxiliary problem 2": 
\begin{equation}
S^{(2)}_{m+1 m} \ldots S^{(2)}_{K m} S^{(2)}_{1 m} \ldots 
S^{(2)}_{m-1 m}
\hat{\Phi}(id) = \prod^{N}_{k=1}\sigma_{k}(a_m)\ \hat{\Phi}(id)
\mb{for} m=1,\ldots,K\,.
\label{AuxPr3 gl21gl2}
\end{equation}

At this stage, we arrived to the third step of nested coordinate
Bethe ansatz. We have only 2 types of possible excitations
$e^{2\uparrow}$ and $e^{3\uparrow}$ and the
$S$-matrix (\ref{S(2)12 gl21gl2}) is a pure permutation.

\subsection{Permutation problem, level three\label{sec:gl21-BAlevel3}}

As mentioned above, we have only 2 types of "particles": $e^{2 
\uparrow}$ and
$e^{3 \uparrow}$ and an equation composed only with permutations:
\begin{equation}
\cH_{j}\,\phi = P_{j+1 j}...P_{K j} P_{1 j}...P_{j-1 j}\, \phi 
= \Lambda\, \phi\,.
\label{PPPphi=Lphi gl21gl2}
\end{equation}

Again, we solve the problem with 
the coordinate Bethe ansatz, and fix the first "particle" $e^{2 
\uparrow}$ 
as the "new" vacuum state
\begin{equation}
\phi_{M=0}=\prod^{K}_{i=1} e^{2 \uparrow}_{i}, \; \; \Lambda=1\,.
\end{equation}
The second "particle" ($e^{3 \uparrow}$) represents some excitations
above this vacuum state.
One can verify that the function
\begin{equation}
\phi_{M}(\vec{n})=\sum_{x_{1}<x_{2}<...<x_{M}} \Psi(\vec{x})
\prod^{M}_{i=1} e^{3 \uparrow}_{x_{i}}
\end{equation}
with
\begin{equation}
\Psi(\vec{x}) = \sum_{P \in S_{M}} \Phi(P) \prod^{M}_{i=1}
g_{x_{i}}(a_{p(i)})\,,\qquad g_{x}(a) = a^x
\end{equation}
is an eigenfunction of this permutation problem if the coefficients 
$\Phi(P)$ obey some conditions that we formulate below.

Acting with the Hamiltonian $\cH_{j}$ gives only the periodicity 
condition on the coefficients $\Phi(P)$:
\begin{equation}
\frac{\Phi(PC)}{\Phi(P)} = [a_{p(M)}]^{K}\,,\quad 
C=\Pi_{1M}...\Pi_{M-1M}\,.
\end{equation}
We assume a simple condition on the coefficients $\Phi(P)$:
\begin{equation}
\Phi(P \Pi_{i i+1}) = \Phi(P) \mb{for}i=1,...,M\,.
\label{eq:ansatzVk}
\end{equation}
Imposing these conditions and periodicity  
leads to
\begin{equation}
g_{x}(a(n)) = e^{\frac{2 \pi i}{K} n x }, \  n=1,...,K\,.
\label{eq:defga}
\end{equation}

Therefore the eigenvector can be written in the form
\begin{equation}
\phi_{M}(\vec{n})=\sum_{x_{1}<x_{2}<...<x_{M}} \sum_{P \in S_{M}}
\prod^{M}_{i=1} e^{\frac{2 \pi i}{K} n_{p(i)} x_{i} } e^{3
\uparrow}_{x_{i}}\mb{with}
1\leq n_{1}<n_{2}<...<n_{M}\leq K\,.
\end{equation}
It has the eigenvalue
\begin{equation}
\Lambda(\vec{n})=\prod^{M}_{j=1} e^{\frac{2 \pi i}{K} n_{j} }
=\exp\left(\frac{2 \pi i}{K}\, |\vec n|\right)\,.
\end{equation}
Then, the total number of states $\phi_{M}(n_{1},...,n_{M})$ is
\begin{equation}
\sum_{0\leq n_{1}<n_{2}<...<n_{M}\leq K-1} 1 = 
\left(\begin{array}{c} K \\ M\end{array}\right) = \frac{K!}{M! (K-M)!}
\end{equation}
which is the right number of eigenfunctions with $M$ excitations. 
Hence the conditions (\ref{eq:ansatzVk}) and the functions 
(\ref{eq:defga}) solve the permutation problem (\ref{PPPphi=Lphi 
gl21gl2}).

Gathering the results of sections \ref{sec:gl21-BAlevel1}, 
\ref{sec:gl21-BAlevel2} and \ref{sec:gl21-BAlevel3}, we get the Bethe 
equations written in section \ref{sec:gl21-BAresu}.

\section{Generalization to $gl(\fn|\fm) \oplus gl(2)$ model\label{sec:glmn}}

In this section we generalize the results obtained in previous 
sections to
 larger algebras, namely the case of $gl(\fn|\fm)_{\uparrow} \oplus 
gl(2)_{\downarrow}$
Hubbard model. The Hamiltonian is given by the expression
\begin{equation}
H_{gl(\fn|\fm) \oplus gl(2)} = \sum_{x=1}^L \left[ (\Sigma P)_{\uparrow \;
x,x+1 } + (\Sigma P)_{\downarrow \; x,x+1 } \right] + u
\sum_{x=1}^L \left( C_{\uparrow \; x} C_{\downarrow \; x} \right),
\label{glNMgl2 Ham}
\end{equation}
with the notation:
\bea
(\Sigma P)_{\uparrow \; x,x+1 } &=& \sum_{a=2}^{\fn+\fm}
\left(
E^{a1}_{\uparrow \; x}E^{1a}_{\uparrow \; x+1} +(-1)^{[a]}
E^{1a}_{\uparrow\; x}E^{a1}_{\uparrow \; x+1} \right) 
\\
(\Sigma P)_{\downarrow \; x,x+1 } &=& E^{12}_{\downarrow \; x}
E^{21}_{\downarrow \; x+1} + E^{21}_{\downarrow \; x}
E^{12}_{\downarrow \; x+1} 
\\
C_{\uparrow\; x} &=&  E^{11}_{\uparrow\; x} - \sum^{\fn+\fm}_{a=2} 
E^{aa}_{\uparrow\; x} 
\quad;\quad
C_{\downarrow\; x} = E^{11}_{\downarrow\; x} -
E^{22}_{\downarrow\; x}
\label{eq:proj-glmn}
\eea
and the grading we use is given in (\ref{eq:grading}).

\subsection{Result for $gl(\fn|\fm) \oplus gl(2)$\label{sec:glnm-BAresu}}

We first gather all results detailed in the following sections and
write down the Bethe equations of our model $gl(\fn|\fm) \oplus gl(2)$:
\bea
&& e^{ik_{j}L}\,=(-1)^{K+N+1} \prod^{K}_{m=1} \frac{i \sin 
k_{j}+ia_{m} +
 \frac{u}{4} }{i \sin k_{j}+ia_{m} - \frac{u}{4}}, \quad  j=1,\ldots,N
\\
&&(-1)^{N} \prod_{j=1}^{N} \frac{i \sin k_{j}+ia_{m} + 
\frac{u}{4} }{i
\sin k_{j}+ia_{m} - \frac{u}{4}} = \Lambda(\vec{n}^{(3)}) 
\prod_{l=1,\; 
l\neq m}^{K} \frac{ ia_{m} - ia_{l} + \frac{u}{2}}{ia_{m} - ia_{l} -
\frac{u}{2}}, \quad  m=1,\ldots,K
\\
&&\Lambda(\vec{n}^{(3)})\, = \exp\left(\frac{2i \pi}{K} |\vec
n^{(3)}|\right), \quad |\vec n^{(3)}|=\sum_{i=1}^{M} n_{i}^{(3)} \nonu
&&0\leq M\leq K \mb{and} 1\leq n^{(3)}_{1}<n^{(3)}_{2}<...<n^{(3)}_{M}\leq
K \eea where $L$ is the number of sites considered in Hubbard model, $N$ is
the total number of $e^{2\downarrow}$,$e^{2\uparrow}$,
$e^{3\uparrow}$,...,$e^{(\fn+\fm)\uparrow}$ "particles". $K$ counts the
total number of excitations from $e^{2\uparrow}$ to $e^{(\fn+\fm)\uparrow}$
and finaly $M$ numbers the $e^{3\uparrow}$,...,$e^{(\fn+\fm)\uparrow}$
"particles".

There are Bethe parameters $n_{i}^{(k)}$, $3< 
k\leq \fm+\fn$, for each particle $e^{k\uparrow}$, but they don't 
show 
up in the Bethe equations.
In section \ref{sec:glnm-BAlevel3}, it is shown more precisely how 
all 
these 
remaining parameters (that are quantized) appear in the Bethe ansatz 
construction.

The energies associated to these states are given by
\begin{equation}
E = (L-2N) + 2 \sum^{N}_{m=1} \cos k_{m} 
\end{equation}
and their momentum reads
\begin{equation}
\fp = \sum^{N}_{m=1} k_{m} \,.
\end{equation}

Let us note that the Bethe equations for $gl(\fn|\fm) \oplus gl(2)$ 
are 
very close to the ones obtained for $gl(2|1) \oplus gl(2)$. This is 
due to the particular projectors we have chosen, see eq. 
(\ref{eq:proj-glmn}). 
More general models (and Bethe equations) can be obtained varying 
these projectors: we come back on this point in section 
\ref{sec:gl22gl2 pi particle}.

\subsection{Coordinate Bethe ansatz, level 1\label{sec:glnm-BAlevel1}}

We solve this model via the coordinate Bethe ansatz. In this model we 
have
$\fn+\fm+2$ types of different "particles" denoted by $e^{1\uparrow},
e^{2\uparrow},..., e^{(\fn+\fm)\uparrow}$ and $e^{1\downarrow},
e^{2\downarrow}$. As in previous case we  choose
the vacuum as
$$
\phi_{0} = \prod^{L}_{x=1} e^{1\uparrow}_{x} e^{1\downarrow}_{x}\,.
$$

The excitations above the vacuum state are given by 
    \begin{equation}
\phi[\overline{A}] = \sum_{\mathbf{x}}
\Psi[\mathbf{x},\overline{A}] e^{A_{1}}_{
x_{1}}...e^{A_{N}}_{x_{N}}
\label{excited st glNM}
    \end{equation}
with indices $A_{j} = 
2\uparrow,3\uparrow,...,(\fn+\fm)\uparrow,2\downarrow$ 
corresponding to vectors
$e^{2\uparrow},e^{3\uparrow}$,...,$e^{(\fn+\fm)\uparrow}$, 
$e^{2\downarrow}$. 
The sum over $\mathbf{x}$ is again considered without points
where two "particles" with the same spin are on the same site. $N$
describes the number of all excitations and goes from 1 to $L$.
In (\ref{excited st glNM}), the sites carrying vectors 
$e^{1\uparrow}_{j} 
e^{1\downarrow}_{j}$, not associated to any excitation, have been 
omitted.

Now we assume the {\itshape Bethe ansatz} for $\Psi(\mathbf{x})$ and follow
the steps detailed in section \ref{sec:gl21}. We divide the coordinate
space $(x_{1},x_{2},..,x_{N})$ into $N!$ sectors. For $x_{q_{1}} <
x_{q_{2}} < .. < x_{q_{N}}$, we have
\begin{equation}
\Psi_{Q}(\mathbf{x}) = \sum_{P} (-1)^{[sg(Q)]} \Phi(P, Q P^{-1}) e^{i 
P\mathbf{k}\cdot
Q\mathbf{x}} \label{Psi ansatz glNM}
\end{equation}
where $P=[p_{1},p_{2},..,p_{N}]$ and $Q=[q_{1},q_{2},..,q_{N}]$ are 
two
permutations of the integers $1,2,..,N$ and  $ 
P\mathbf{k}\cdot
Q\mathbf{x} = \sum_{i} k_{p_{i}} x_{q_{i}}$.  
As in previous section, the symbol $(-1)^{[sg(Q)]}$ stands for
 the signature of the $Q$-permutation when restricted to fermionic particles
 $e^{(\fn +1)\uparrow}$,...,
$e^{(\fn +\fm)\uparrow}$. We recall that
$\Psi(\mathbf{x})$ and accordingly $\Phi(P,QP^{-1})$ both depend on 
the type of
excitations $\overline{A}$.

We gather all the coefficients $\Phi[P, Q
P^{-1},\overline{A}]$ in a vector
$$
\hat{\Phi}(P) \equiv \sum_{Q,\overline{A}}
\Phi[P,QP^{-1},\overline{A}]
|A_{q_{1}},..,A_{q_{N}}>
$$
where the sum is done over all possible types $\overline{A}$ 
and all
corresponding sectors $Q$. A vector
$|A_{q_{1}},...,A_{q_{N}}> $ represents
one state of $N$ possible excitations and belongs to $V_{1} \otimes 
...
\otimes V_{N}$ (where $V=\spn\{2\downarrow;
2\uparrow;3\uparrow;4\uparrow;...;(\fn+\fm)\uparrow
\}$).

In the case of only two particles ($N=2$) we are able to introduce the
 $S$-matrix:
$$
\hat{\Phi}(\Pi_{12}P) = S^{(1)}_{12}(\lambda_{1}-\lambda_{2}) 
\hat{\Phi}(P)\,.
$$
Here, $S^{(1)}_{12}(\lambda_{1}-\lambda_{2})\equiv S^{(1)}_{12}$ acts 
on elementary vectors
$|A_{1},A_{2}>$ as
\bea
S^{(1)}_{12} |2\downarrow,2\downarrow> &=& -
|2\downarrow,2\downarrow>, \\
S^{(1)}_{12} |2\downarrow,A> &=& t_{12}\,
|2\downarrow,A> + r_{12}\, |A,2\downarrow>, \\
S^{(1)}_{12} |A,B> &=& - |B,A>,
\mb{for}A,B = 2\uparrow,...,(\fn+\fm)\uparrow
\eea
 where
$$
t_{12} = \frac{2i(\lambda_{1} - \lambda_{2} )}{u + 2i (\lambda_{1} -
\lambda_{2}) }, \; \; r_{12} = \frac{ - u }{u + 2i (\lambda_{1} -
\lambda_{2}) }\,.
$$

For an arbitrary number of excitation $N$ we have
$$
\hat{\Phi}(\Pi_{ab}P) = S^{(1)}_{ab}(\lambda_{a}-\lambda_{b})\,
\hat{\Phi}(P),
$$
where the matrix $S^{(1)}_{ab}$ acts nontrivialy only on $V_{a} 
\otimes V_{b}$ 
vector
spaces.

Again, the periodic boundary condition on the function
$\Psi(\mathbf{x})$
$$
\Psi_{QC}(\mathbf{x}+\mathbf{e}_{q_{1}}L)=\Psi_{Q}(\mathbf{x})
\mb{with}C=\Pi_{N 1}...\Pi_{N N-1}
$$
leads to the second step in our problem: 
\begin{equation}
S^{(1)}_{j+1 j}...S^{(1)}_{N j} S^{(1)}_{1 j}...S^{(1)}_{j-1 j}
\hat{\Phi}(id) = e^{ik_{j}L} \hat{\Phi}(id) 
\qquad j=1,\ldots,N\,.
\label{AuxPr1 glNMgl2}
\end{equation}

\subsection{Auxiliary problem, level 2\label{sec:glnm-BAlevel2}}

As in the previous section, we transform slightly the $S$-matrix
$S^{(1)}_{12}\rightarrow - S^{(1)}_{12} \equiv S_{12}$ to simplify the
calculations. The idea is that we work in different steps to 
diagonalize
these matrices $S_{12}$ via the coordinate Bethe ansatz. At each step 
we
specify a "new" vacuum and "new" excitations. At this step we have 
$\fn+\fm$
types of different excitations: $e^{2 \downarrow},e^{2 \uparrow}, e^{3
\uparrow},\ldots,e^{(\fn+\fm) \uparrow}$.

Choosing the "new" vacuum as
\begin{equation}
\phi_{0} = \prod^{N}_{k=1} e^{2 \downarrow}_{k},
\end{equation}
the state of $K$ excitations of any type is written as
\begin{equation}
\phi(\bar{A}) = \sum_{\mathbf{x}} \Psi(\mathbf{x}, \bar{A})
e^{A_{1}}_{x_{1}}...e^{A_{K}}_{x_{K}}
\label{eq:excitLevel2}
\end{equation}
where the sum is done over all coordinates $x_{i}$ without coinciding
points $x_{i}=x_{k}$ for any $i,k$ (exclusion principle). $\bar{A}$ is a
vector $(A_{1},..,A_{K})$ with $A_{i}=2 \uparrow, 3 \uparrow, ...,
(\fn+\fm) \uparrow$ (corresponding to $e^{2 \uparrow}, e^{3 \uparrow}$
...$e^{(\fn+\fm) \uparrow}$). Again, sites carrying $e^{2\downarrow}$ (no
excitation) have been omitted in (\ref{eq:excitLevel2}).

The Bethe ansatz for the coefficients $\Psi(\mathbf{x},\bar{A})$, in 
the
sector $x_{q_{1}}<x_{q_{2}}<...<x_{q_{K}}$ where 
$Q=[q_{1},...,q_{K}]$ is
the permutation of the integers $1,2,...,K$, is given by
\begin{equation}
\Psi_{Q}(\mathbf{x},\bar{A}) = \sum_{P} \Phi(P,QP^{-1}, \bar{A})
f_{x_{q_{1}}}(a_{p_{1}}) f_{x_{q_{2}}}(a_{p_{2}})
...f_{x_{q_{K}}}(a_{p_{K}})\,.
\end{equation}
 $f_{x}(a)$ is the one-particle solution (for 
any type of excitation $A$) with rapidity $a$:
\begin{equation}
f_{x}(a) = \prod^{x-1}_{m=1} \left(-\frac{i \lambda_{m} + i a +
\frac{u}{4}}{i \lambda_{m+1} + i a - \frac{u}{4}} \right).
\label{f glNM}
\end{equation}

The eigenvalue corresponding to this state $\phi(\bar{A})$
takes the form
\begin{equation}
\Lambda_{j} = \sigma_{j}(a_1)...\sigma_{j}(a_K)
\end{equation}
where $\sigma_{j}(a)$ is the eigenvalue of the one-particle solution
\begin{equation}
\sigma_{j}(a) = - \frac{i \lambda_{j} + i a + \frac{u}{4}}{i 
\lambda_{j} +i a - \frac{u}{4}}\,.
 \label{sigma glNM}
\end{equation}

Next we gather all the coefficients $\Phi(P,QP^{-1}, \bar{A})$ in a
vector
\begin{equation}
\hat{\Phi}(P) \equiv \sum_{Q,\overline{A}} 
\Phi(P,QP^{-1},\overline{A})
|A_{q_{1}},..,A_{q_{K}}>
\end{equation}
where the sum is over all types of excitations and all 
corresponding
sectors. The vector $|A_{q_{1}} ,..,A_{q_{K}}> $ belongs to $V_{1} 
\otimes ...
\otimes V_{K}$ (where $V=\spn\{e^{2 \uparrow};e^{3 
\uparrow};...;e^{(\fn+\fm)
\uparrow}\}$) and represents one combination of $K$ excitations.

Application of the Hamiltonian (\ref{AuxPr1 glNMgl2}) on the excited 
state
$\phi(\bar{A})$ gives two types of conditions imposed on the 
coefficients
$\hat{\Phi}(P)$. From the first one, in the case of only two "particles"
($K=2$), we can introduce again a $S$-matrix corresponding to 
the
second step. The second type implies the periodicity condition. Thus,
the $S$-matrix is defined by
\begin{equation}
\hat{\Phi}(\Pi_{12}P) = S^{(2)}_{12}(a_{1}-a_{2}) \hat{\Phi}(P)
\end{equation}
with
\begin{equation}
S^{(2)}_{12}(a_{1}-a_{2}) = \alpha_{12}P_{12}, \; \; \text{and} \; \;
\alpha_{12} =  \frac{ia_{1} - ia_{2} + \frac{u}{2}}{ia_{1} - ia_{2} -
\frac{u}{2}}\,.
\end{equation}

For an arbitrary number $K$ of excitations we have
\begin{equation}
\hat{\Phi}(\Pi_{ij}P) = S^{(2)}_{ij}(a_{i}-a_{j}) \hat{\Phi}(P),
\end{equation}
where $p^{-1}(j)-p^{-1}(i)=1$ and the permutation 
$S^{(2)}_{ij}(a_{i}-a_{j})$
acts nontrivialy only on $V_{i} \otimes V_{j}$ vector spaces. 
The matrix $S^{(2)}_{12}(a_{1}-a_{2})$ 
satisfies the Yang--Baxter equation since it is a permutation.

The periodic boundary conditions on $\hat{\Phi}(P)$ implied by the 
action
of the Hamiltonian (\ref{AuxPr1 glNMgl2}) is written in following 
form:
\begin{equation}
S^{(2)}_{m+1 m}...S^{(2)}_{K m} S^{(2)}_{1 m}...S^{(2)}_{m-1 m}
\hat{\Phi}(id) = \prod^{N}_{k=1}\sigma_{k}(a_m) \hat{\Phi}(id)
\mb{for} m=1,...,K\,,
\label{AuxPr3 glNM}
\end{equation}
where the $S$-matrix arguments were omitted for simplicity.

\subsection{Permutation problem, level 3\label{sec:glnm-BAlevel3}}

Thus, we arrive to the third step of nested coordinate Bethe ansatz. 
 Here, we have $e^{2 \uparrow}$, $e^{3 \uparrow}$, ... ,
$e^{(\fn+\fm )\uparrow}$ "particles" that move `freely', the 
Hamiltonian $\Gamma$
being constructed on permutations only:
\begin{equation}
\Gamma \,\phi=P_{j+1 j}...P_{K j} P_{1 j}...P_{j-1 j}\ \phi = \Lambda\, 
\phi\,.
\label{PPPphi=Lphi glNMgl2}
\end{equation}
Note that $\Gamma$ is a cyclic permutation, and is independent from 
$j$.

We choose  the "particle" 
$e^{2\uparrow}$ as the vacuum state:
\begin{equation}
\phi_{M=0}=\prod^{K}_{i=1} e^{2 \uparrow}_{i}\,, \quad  \Lambda=1\,,
\end{equation}
and introduce the function
\begin{equation}
\phi^{(3)}_{M}(\bar{A})=\sum_{\vec{x}} \Psi(\vec{x}) \prod^{M}_{i=1}
e^{A_{i}}_{x_{i}}, \;\; A_{i}=3 \uparrow, 4 \uparrow,..., (\fn+\fm) 
\uparrow\,.
\label{phi3}
\end{equation}
 It describes a state with $M$ excitations above the vacuum state 
$\phi_{M=0}$. 

The coefficients $\Psi(\vec{x})$ are defined in the sector $1\leq 
x_{q(1)}<x_{q(2)}<...<x_{q(M)}
\leq K$, with $Q \in \fS_{M}$,  by
\begin{equation}
\Psi_{Q}(\vec{x}) = \sum_{P \in \fS_{M}} \Phi^{(3)}(P,QP^{-1})
\prod^{M}_{i=1} g_{x_{q(i)}}(a^{(3)}_{p(i)})\,,\qquad g_{x}(a) = 
a^x\,.
\label{psi3}
\end{equation}
One can verify that $\phi^{(3)}_{M}(\vec{A})$ is an eigenfunction
 with the following eigenvalue
\begin{equation}
\Lambda = \prod_{i=1}^{M} a^{(3)}_{i}
\end{equation}
if some conditions, which we precise below, are satisfied.

Application of the Hamiltonian gives only the periodicity condition 
on 
the coefficients $\Phi^{(3)}(P,QP^{-1})$:
\begin{equation}
\frac{\Phi^{(3)}(PC,QP^{-1})}{\Phi^{(3)}(P,QP^{-1})} =
[a^{(3)}_{p(M)}]^{K}\,,\qquad C=\Pi_{1M}...\Pi_{M-1M}\,.
\end{equation}
As in section \ref{sec:gl21-BAlevel3}, we assume some relations
on the coefficients
$\Phi^{(3)}(P,QP^{-1})$, but, now, the form of these relations depend 
on 
whether the particles are identical or not. 
If, for a given $i$, $x_{i}$ and $x_{i+1}$ correspond to identical 
particles 
we impose
\begin{equation}
\Phi^{(3)}(\Pi_{p(i) p(i+1) } P ,QP^{-1}) = 
\Phi^{(3)}(P,QP^{-1}),
\label{PhiPabP=PhiP1}
\end{equation}
while, otherwise, we set
\begin{equation}
\Phi^{(3)}(\Pi_{p(i) p(i+1) } P ,QP^{-1} \Pi_{p(i) p(i+1) 
} ) =
\Phi^{(3)}(P,QP^{-1})\,.
\label{PhiPabP=PhiP2}
\end{equation}

As we can see, there is a sector changing in the relations 
above, and we proceed recursively using the same methods as above. 
We introduce 
\begin{equation}
\hat{\Phi}^{(3)}(P) \equiv \sum_{Q,\bar{A}}
\Phi^{(3)}[P,Q,\bar{A}] |A_{q_{1}},..,A_{q_{M}}>
\label{Phi^{(3)}}
\end{equation}
where the sum is over possible types $\bar{A}$ and all
corresponding sectors $Q \in \fS_{M}$. The vector 
$|A_{q_{1}},...,A_{q_{N}}>$ 
represents
one state with $M$ excitations and belongs to $V_{1} \otimes 
...
\otimes V_{N}$ (where $V=\spn\{e^{3 \uparrow}, e^{4 \uparrow}, ... , 
e^{(\fn+\fm)\uparrow} \}$).
Then, relations (\ref{PhiPabP=PhiP1}) and (\ref{PhiPabP=PhiP2}) can 
be rewritten in the following form
\begin{equation}
\hat{\Phi}^{(3)}(\Pi_{ab} P) = S^{(3)}_{ab} 
\hat{\Phi}^{(3)}(P)\,,\qquad S^{(3)}_{ab}=P_{ab}\,.
\label{PhiPabP=Sab PhiP}
\end{equation}

Therefore, the periodic boundary conditions on $\hat{\Phi}(P)$ 
implied by
the action of the chain of permutations is written as
\begin{equation}
P_{m+1 m}...P_{M m} P_{1 m}...P_{m-1 m} \hat{\Phi}^{(3)}(id) =
[a^{(3)}_{m}]^{K} \hat{\Phi}^{(3)}(id)\,,\quad m=1,...,M
\end{equation}
but here we have already only $e^{3 \uparrow}$, ... 
,$e^{(\fn+\fm)\uparrow}$
"particles" involved in the calculations. Thus, we arrive to the 
next level of nested coordinate Bethe ansatz, with, again, an 
Hamiltonian 
built on permutations only, and a new chain of length $M$.

Using the previous considerations, we repeat the same method and we
"eliminate"  one by one the "particles" $e^{3 \uparrow}$, ... up to
$e^{(\fn+\fm-1 )\uparrow}$, choosing it as the vacuum state at each 
nested level. 

We suppose that we have $M_{3}$ "particles" of type $e^{3\uparrow}$,
$M_{4}$ of type $e^{4 \uparrow}$,..., $M_{\fn+\fm}$ of type $e^{(\fn+\fm)
\uparrow}$, so that $M_{3}+ M_{4}+...+M_{\fn+\fm}=M$. At each level
$k=3,...,\fn+\fm-1$, we
have particles $e^{(k+1)\uparrow}, ..., e^{(\fn+\fm) \uparrow}$ as 
different types of 
excitations above the vacuum state built on $e^{k \uparrow}$.
The eigenvector $\hat{\Phi}^{(k)}(id)$ can be written in the same 
form as in (\ref{phi3}) and (\ref{psi3}), with the set of Bethe roots 
$\{ a^{(k+1)}_{i} \}_{i=1}^{M_{k+1}+...+M_{\fn+\fm}}$ and the 
coefficients $\Phi^{(k+1)}(P,QP^{-1})$ with $P,Q \in 
\fS_{M_{k+1}+...+M_{\fn+\fm}}$. These coefficients are used 
 to write the vector $\hat{\Phi}^{(k+1)}(id)$ 
% 
% Similarly, the eigenvalue will be
% \begin{equation}
% \Lambda^{(k)} = \prod_{i=1}^{M_{k+1}+...+M_{\fn+\fm}} a^{(k+1)}_{i} 
% \end{equation}
% and 
that obeys the periodicity condition:
\begin{equation}
P_{m+1, m}...P_{M_{k+1}+...+M_{\fn+\fm}, m} P_{1 m}...P_{m-1, m} 
\hat{\Phi}^{(k+1)}(id) =
\left[ a^{(k+1)}_{m} \right]^{M_{k}+...+M_{\fn+\fm}} 
\hat{\Phi}^{(k+1)}(id),
\end{equation}
for $m=1,...,(M_{k+1}+...+M_{\fn+\fm})$. 

Also, we find the periodicity 
condition of the previous level  (when we pass from level $k-1$ to level $k$)
\begin{equation}
\left[ a^{(k)}_{m} \right]^{M_{k-1}+...+M_{\fn+\fm}} = 
\prod_{i=1}^{M_{k+1}+...+M_{\fn+\fm}} a^{(k+1)}_{i}  
\label{permutation rec Bae}
\end{equation}
for $m=1,...,(M_{k+1}+...+M_{\fn+\fm})$ and 
$k=3,...,\fn+\fm-1$ with $M_{2} = K - (M_{3} +...+M_{\fn+\fm})$.

At last level, we have only one type of excitations $e^{(\fn+\fm) 
\uparrow}$ on the vacuum state $e^{(\fn+\fm - 1)\uparrow}$ and we can 
see that it is the same case as in the permutation problem of the 
model $gl(2|1) \oplus gl(2)$. Thus, using the relation 
(\ref{permutation rec Bae}) for $k=\fn+\fm-1$, we find the following 
Bethe equations which link the case of one type excitation 
$e^{(\fn+\fm)\uparrow}$ and the previous level with two types of 
excitations $\{ e^{(\fn+\fm-1)\uparrow},e^{(\fn+\fm)\uparrow} \}$:
\begin{equation}
[a^{(\fn+\fm-1)}_{i}]^{M_{\fn+\fm-2}+M_{\fn+\fm-1}+M_{\fn+\fm}}= 
\prod_{j=1}^{M_{\fn+\fm}}
a^{(\fn+\fm)}_{j},\quad i=1,\ldots,M_{\fn+\fm-1}+M_{\fn+\fm}\,,
\end{equation}
together with the result obtained from $gl(2|1) \oplus gl(2)$ model
\bea
&&a^{(\fn+\fm)}_{j} = e^{2 \pi i 
\frac{n^{(\fn+\fm)}_{j}}{M_{\fn+\fm-1}+M_{\fn+\fm}}},
\quad j=1,\ldots,M_{\fn+\fm}\\
&& 1 \leq  
n_{1}^{(\fn+\fm)}<n_{2}^{(\fn+\fm)}<...<n_{M_{\fn+\fm}}^{(\fn+\fm)}  
\leq M_{\fn+\fm-1}+M_{\fn+\fm}\,. 
\eea

In the same way, we can write the Bethe equations corresponding to the 
transition between level with $\{ e^{(\fn+\fm-1) \uparrow}, 
e^{(\fn+\fm)\uparrow} \}$ and the previous 
one
\begin{equation}
[a^{(\fn+\fm-2)}_{i}]^{M_{\fn+\fm-3}+...+M_{\fn+\fm}}=
\prod_{j=1}^{M_{\fn+\fm-1}+M_{\fn+\fm}} a^{(\fn+\fm-1)}_{j},\;\;
i=1,\ldots,M_{\fn+\fm-2}+...+M_{\fn+\fm}
\end{equation}
and we can continue this recurrence up to $a^{(3)}$.

Solutions for every set of Bethe roots $a^{(k)}$ can be computed as 
following. For $k=3,...,\fn+\fm-1$ with $M_{2} = K - (M_{3} 
+...+M_{\fn+\fm})$ we have
\bea
&&a^{(k)}_{j}= e^{ \frac{2 \pi i }{M_{k-1}+...+M_{\fn+\fm}}  \left[ 
n^{(k)}_{j} + \frac{\sum_{m=k+1}^{\fn+\fm} \sum_{i=1}^{M_{m}+...+M_{\fn+\fm}} 
n^{(m)}_{i} }{M_{k}+...+M_{\fn+\fm}} 
\right]  }\,,\quad
j=1,\ldots,(M_{k}+...+M_{\fn+\fm})\quad\nonu
&& 1 \leq  n^{(k)}_{1}<n^{(k)}_{2}<...<n^{(k)}_{M_{k}+...+M_{\fn+\fm}} 
\leq  M_{k-1}+...+M_{\fn+\fm} \,.
\eea

Therefore, the function is the eigenvector of the initial permutation 
problem
(\ref{PPPphi=Lphi glNMgl2})
\begin{equation}
\phi^{(3)}_{M}(\bar{A})=\sum_{Q \in \fS_{M}} \sum_{\vec{x} \in Q} 
\sum_{P
\in \fS_{M}} \Phi^{(3)}(P,QP^{-1}) \prod^{M}_{i=1}
[a^{(3)}_{p(i)}]^{x_{q(i)}} e^{A_{i}}_{x_{i}}\ \phi_{0}(2\uparrow)
\end{equation}
where we denoted explicitely the `empty sites' (with no excitation) 
as 
$\phi_{0}(2\uparrow)$. 

The coefficients $\Phi^{(3)}(id,Q)$ gathered in $\hat{\Phi}^{(3)}(id)$ are
connected with the next level coefficients $\Phi^{(4)}(id,Q)$. To see it,
we fix the value of $M_{3}$ and consider vectors $\bar{B}$ which
characterizes excitations of the form
\begin{equation}
\bar{B}=(\overbrace{(\fn+\fm) \uparrow,...,(\fn+\fm) 
\uparrow,...,4\uparrow,...,4\uparrow}^{M-M_{3}},
\overbrace{3\uparrow,...,3\uparrow}^{M_{3}})\,.
\end{equation}
Then, we consider the restriction of the relation (\ref{Phi^{(3)}}) to 
\begin{equation}
\left.\hat{\Phi}^{(3)}(id)\right|_{restric.} =  \sum_{Q,\bar B}
\Phi^{(3)}[id,Q,\bar{B}] |B_{q_{1}},..,B_{q_{M}}> \,.
\end{equation}
One can recognize in this term the $\phi^{(4)}_{M-M_{3}}(\bar{B})$ 
coefficient
(here we take only first $M-M_{3}$ values $\bar{B}$):
\begin{equation}
\sum_{Q}
\Phi^{(3)}[id,Q,\bar{B}] 
e^{B_{1}}_{q^{-1}(1)}...e^{B_{M-M_{3}}}_{q^{-1}(M-M_{3})}
\phi_{0}(3 \uparrow)
% |(3 \uparrow)_{1},..,(3 \uparrow)_{M}> = 
= \sum_{\vec{y}\in [1,M]}
\Psi(\vec{y}) \prod_{i=1}^{M-M_{3}} e^{B_{i}}_{y_{i}} \phi_{0}(3 
\uparrow) \equiv  \phi^{(4)}_{M-M_{3}}(\bar{B}), 
\end{equation}
with $y_{i}=q^{-1}(i)$ for $i=1,...,M-M_{3}$. In the left hand side 
of the equation $e^{B_{i}}_{q^{-1}(i)}$ are the operators which 
create the corresponding excitations $B_{i}$ on the site $q^{-1}(i)$ 
of the chain of particles $e^{3 \uparrow}$.  

Therefore, using the same ansatz as in (\ref{psi3}) for 
$\Psi(\vec{y})$ in $\phi^{(4)}_{M-M_{3}}(\bar{B})$ we can identify 
the coefficients $\Phi^{(3)}(id,Q)$ as
\begin{equation}
\Phi^{(3)}(id,Q, \bar{B}) = \sum_{P \in
\fS_{M-M_{3}}} \Phi^{(4)}(P,Q'P^{-1},\bar{B}) 
\prod^{M-M_{3}}_{i=1}
[a^{(4)}_{p(i)}]^{i},
\label{phi3 prop phi4}
\end{equation}
with $Q' \in \fS_{M-M_{3}}$ defined by $q'(i)=q(i)$ for 
$i=1,...,M-M_{3}$ and 
\begin{equation}
B_{i}= 4 \uparrow,..., (\fn+\fm) \uparrow,\;\; i=1,...,M-M_{3}\,.
\end{equation}

In the general case, the coefficients $\Phi^{(k)}(P,Q)$ are defined by the 
same
relations: for $k=3,...,\fn+\fm -2$, we have
\begin{equation}
\Phi^{(k)}(id,Q,\bar{B}) = \sum_{P \in \fS_{M_{k+1}+...+M_{\fn+\fm}}} 
\Phi^{(k+1)}(P,Q'P^{-1}, \bar{B}) 
\prod^{M_{k+1}+...+M_{\fn+\fm}}_{i=1} [a^{(k+1)}_{p(i)}]^{i},
\label{phik prop phik+1}
\end{equation} 
where $Q' \in \fS_{M_{k+1}+...+M_{\fn+\fm}}$ is defined by 
$q'(i)=q(i)$ for $i=1,...,M_{k+1}+...+M_{\fn+\fm}$,
$$
B_{i}= (k+1)\uparrow,..., (\fn+\fm) \uparrow,\;\;
i=1,...,M_{k+1}+...+M_{\fn+\fm}
$$
and there are relations similar to (\ref{PhiPabP=PhiP1}) and
(\ref{PhiPabP=PhiP2}).

At last, when $k=\fn+\fm -1$, using the results of $gl(2|1) \oplus 
gl(2)$ model, we get
\begin{equation}
\Phi^{(\fn+\fm -1)}(id,Q,\bar{B}) = 
\sum_{P \in \fS_{M_{\fn+\fm -1}+M_{\fn+\fm}}} \prod^{M_{\fn+\fm 
-1}+M_{\fn+\fm}}_{i=1} [a^{(\fn+\fm)}_{p(i)}]^{i}.
\end{equation}
Equations of the type (\ref{phi3 prop phi4}) and (\ref{phik prop 
phik+1}) together with relations 
(\ref{PhiPabP=PhiP1}) and (\ref{PhiPabP=PhiP2}) 
allow us to derive all the coefficients $\Phi^{(3)}(P,Q)$.
The eigenvalue  reads
\begin{equation}
\Lambda(\vec{n}^{(3)},...,\vec{n}^{(\fn+\fm)})
\equiv \Lambda(\vec{n}^{(3)})= \prod_{i=1}^{M} 
a^{(3)}_{i} =
\exp\Big(\frac{2 \pi i}{K} \sum^{M}_{i=1} n^{(3)}_{i}\Big)
=
\exp\Big(\frac{2 \pi i}{K} |\vec n^{(3)}|\Big)\,.
\end{equation}
The Bethe parameters $\vec n^{(k)}$, $k>3$, ensure the correct 
multiplicity of eigenfunctions.
Indeed, the total number of states
$\phi_{M}(\vec{n}^{(3)},\vec{n}^{(4)},...,\vec{n}^{(\fn+\fm)})$ is
\bea
\sum_{\atopn{\vec{n}^{(3)},\vec{n}^{(4)},
...,\vec{n}^{(\fn+\fm)}}{1\leq 
n^{(k)}_{i}<n^{(k)}_{i+1}\leq J_{k}}} 1 &=&
\frac{K!}{M_{3}!...M_{\fn+\fm}! (K-M)!},
\mb{where} J_{k}=\sum_{\ell=3}^{\fm+\fn-k-1}M_{\ell},
\eea
which shows that the ansatz is complete.

\section{Another $gl(2|2) \oplus gl(2)$ model\label{sec:gl22gl2 pi particle}}

\subsection{Comparison with AdS/CFT models \label{sec:AdScft}}
It is known that the Hubbard model can be connected to the $SU(2)$ 
subsector of the super-Yang--Mills (SYM) theory. The dilatation 
operator in 
this subsector can be identified with a Hamiltonian that is very 
close to the Hubbard one. Although not exact\footnote{The 
correspondence breaks down at level 4 of perturbation theory 
\cite{Rej:2005qt}}, this 
correspondence has shed a new light on the integrability aspect of 
SYM models. In fact, one can introduce perturbatively a scattering matrix 
(obeying the
Yang--Baxer equation) that differs from the Hubbard one by a phase (so-called 
`wrapping problem'). This phase is
also seen in the Bethe equation of the model. 

The models we have presented up to now possess the same property: they have 
Bethe equations that are very close to the Hubbard Bethe equations, 
but a phase. Unfortunately, this phase is built on `hidden' Bethe 
parameters, but 
not on the impulsions of our `particles' (which is the case of SYM 
Bethe equations). Hence, the correspondence is not immediate, but the 
present construction gives a way to introduce a phase in the 
equations. To strengthen the present approach, 
we show in this section an example of another $gl(2|2) \oplus gl(2)$ model 
using different choices of the projectors $\pi$ and $\bar\pi$. It 
will lead to Bethe equation with a phase that (partially) depends on 
the impulsions of the particles.

We recall that the Hamiltonian is given by 
   \begin{equation}
H_{gl(2|2) \oplus gl(2)} = \sum_{x=1}^{L} \Big( (\Sigma P)_{\uparrow 
\;
x,x+1 } + (\Sigma P)_{\downarrow \; x,x+1 }  + u \ 
C_{\uparrow x} \, C_{\downarrow x} \Big),
\label{gl22gl2 Ham}
    \end{equation}
where we  choose the projectors $\pi_\uparrow$ and 
$\pi_\downarrow$ such that $\cN_{\uparrow} = \{1,2\}$ and 
$\overline\cN_{\uparrow} = \{3,4\}$. Thus
    \begin{eqnarray}
&&(\Sigma P)_{\uparrow \; x,x+1 } = \sum_{i=1}^2 
\sum_{j=3}^4 \Big( E^{ij}_{\uparrow \; x}
E^{ji}_{\uparrow \; x+1} + E^{ji}_{\uparrow \; x}
E^{ij}_{\uparrow \; x+1} \Big)\,,
\\
&& (\Sigma P)_{\downarrow \; x,x+1 } = E^{12}_{\downarrow \; x}
E^{21}_{\downarrow \; x+1} + E^{21}_{\downarrow \; x}
E^{12}_{\downarrow \; x+1}\,, \\
&& C_{\uparrow\; x} = E^{11}_{\uparrow\;x}
+E^{22}_{\uparrow\;x} - E^{33}_{\uparrow\; x} - E^{44}_{\uparrow\; x} 
\mb{;}
 C_{\downarrow\; x} = E^{11}_{\downarrow\; x} - 
E^{22}_{\downarrow\; x}\,.
\end{eqnarray}

We use the same approach, i.e. the coordinate Bethe ansatz, to solve this 
model. Here we do not give all explicit details of calculation. 
The method is the same, however there are some modifications appearing 
when we pass from the initial problem to the first auxiliary problem 
and then to the second auxiliary problem. We briefly give the most 
important statements as well as the Bethe equations.       

\subsection{Bethe equations for $gl(2|2) \oplus gl(2)$\label{sec:Bethepi}}
    
As we shall see in the next subsection for construction of the Bethe 
ansatz, in this model
there are four different kinds of "particles" above the vacuum state: 
$e^{2\uparrow}$ which is defined by projectors above as 
$\pi$-particle and 
$e^{2\downarrow}$,$e^{3\uparrow}$ and $e^{4\uparrow}$ which are the 
$\bar\pi$-particles. To define the Bethe equations, we first 
introduce 
\beq
\bA=\{a_{1},a_{2},\ldots,a_{N_{1}}\} \mb{for some integers such that}
1 \leq a_{1} < a_{2} <...< a_{N_{1}} \leq N\,.
\eeq
Then, the Bethe equations can be written as
\bea
&& e^{ik_{j}(L-N_{2}-N_{3})}=(-1)^{N_{1}}
\mb{for} j\in\bA
\\ 
&& e^{ik_{j}L}=
(-1)^{N+1-(N_{1}+N_{2}+N_{3})} \prod^{N_{2}+N_{3}}_{m=1} \frac{i \sin 
k_{j}+ib_{m} + \frac{u}{4} }{i \sin k_{j}+ib_{m} - \frac{u}{4}} \;,
\mb{for} j\in [1,N]\setminus\bA
\\
&&(-1)^{N-N_{1}} \prod_{\atopn{j = 1}{j \not\in \bA} }^{N} 
\frac{i \sin k_{j}+ib_{m} + 
\frac{u}{4} }{i\sin k_{j}+ib_{m} - \frac{u}{4}} = 
\Lambda(\vec{n})  \prod_{j\in\bA} e^{- i k_{j}} 
\prod_{\atopn{l = 1}{l \neq m}}^{N_{2}+N_{3}} \frac{ 
ib_{m} - ib_{l} + \frac{u}{2}}{ib_{m} - ib_{l} - 
\frac{u}{2}}\;,\quad \label{eq:BEb}\\
&&\mb{for} m=1,\ldots,N_{2}+N_{3} \nonu
&&\Lambda(\vec{n}) = \exp\left(\frac{2i \pi}{N_{2}+N_{3}} \sum_{i=1}^{N_{3}} 
n_{i}\right),\;\; 1\leq n_{1}<n_{2}< \ldots <n_{N_{3}} \leq N_{2}+N_{3} 
\eea
where $L$ is the number of sites considered in Hubbard model, $N$ is 
total
number of all $e^{2\uparrow}$, $e^{2\downarrow}$,$e^{3\uparrow}$ and 
$e^{4\uparrow}$  "particles". $N_{1}$ counts $e^{2\uparrow}$ 
excitations, $N_{2},N_{3}$ count respectively $e^{3\uparrow}$ and 
$e^{4\uparrow}$ particles. Remark that with respect to the  
 Bethe equations computed in the previous sections, the phase 
 $\Lambda(\vec n)$ has been changed to
$$
\Lambda(\vec n)\quad\to\quad \Lambda(\vec n)\prod_{j\in\bA} e^{- i k_{j}} 
$$
showing a (partial) dependence on the momenta of the particles. This 
"dressing" of the phase is similar to the one suggested in 
\cite{dressing}.

The energy associated to the state is given by
\beq
E  = L-2(N-N_{1}) +2 \sum_{j\in[1,N]\setminus\bA} \cos(k_j) 
\eeq
and the momentum reads
\beq
\fp = \sum_{j=1}^N k_{j}.
\eeq
The set of integers $\bA$ is related to the 
$\pi$-particles in the first auxiliary problem and the integers $n_{j}$ correspond 
to the Bethe parameters of the last level, but their Bethe equations 
have already been solved: they just correspond to 
the quantization of these parameters. The parameters 
$\{b_{l}\}_{l=1,\ldots,N_{2}+N_{3}}\equiv 
\{a_{N_{1}+l}\}_{l=1,\ldots,N_{2}+N_{3}}$ correspond to 
$\bar\pi$-particles in the first auxiliary problem, and do have Bethe 
equations, see eq. (\ref{eq:BEb}).

\subsection{Calculation description for $gl(2|2) \oplus gl(2)$}        
    
In this paragraph we briefly describe some important points of the 
approach for this new model. 
At the first level of coordinate Bethe ansatz, together with 
the $\bar\pi$-particles (the "physics" of which we studied 
above), we include some 
$\pi$-particles. The states of $N$ excitations can be written as
    \begin{equation}
\phi[\overline{A}] = \sum_{\mathbf{x}}
\Psi[\mathbf{x},\overline{A}]
e^{A_{1}}_{x_{1}}...e^{A_{N}}_{x_{N}}
\label{excited st gl22gl2}
    \end{equation} 
with $A 
=(2,\uparrow)\,;\,(3,\uparrow)\,;\,(4,\uparrow)\,;\,(2,\downarrow)$.

This modifies the ansatz for the wave function as, for $x_{q_{1}} < 
x_{q_{2}} < \ldots < x_{q_{N}}$
\begin{equation}
\Psi^{P_{\pi\bar\pi}}_{Q}(\mathbf{x}) = \sum_{P'=P_{\pi}P_{\bar\pi}} 
\Phi(\hat{P} Q, \hat{P}^{-1}) e^{i \hat{P}\mathbf{k}\cdot
\mathbf{x}} \label{Psi ansatz gl22gl2},\;\;\hat{P}=P_{\pi\bar\pi}P'
\end{equation}
with the energy:
\begin{equation}
E^{P_{\pi\bar\pi}} = 2 \sum_{l\in\bar\pi} \cos(k_{P_{\pi\bar\pi}(l)}) 
+ L-2(N-N_{1})  \label{free energie gl22gl2}
\end{equation}
where $Q \in \fS_N$. We have to consider the permutation of Bethe 
roots $k_{j}$ in some factorized form: $P'=P_{\pi}P_{\bar\pi}$ 
where the terms permute only $\pi$ and $\bar\pi$ particles 
separately. In addition, we could also vary the value of energy by 
mixing the impulsions of all particles, adding a permutation 
$P_{\pi\bar\pi}$ in the term $e^{i P\mathbf{k}\cdot \mathbf{x}}$, but it 
does not produce new (independent) eigenvectors. 

Applying the Hamiltonian (\ref{gl22gl2 Ham}) on the vector 
(\ref{excited st gl22gl2}) we find the relations between the 
coefficients $\Phi(\hat{P} Q, \hat{P}^{-1})$. Again we can gather all 
relations in a vector $\hat{\Phi}(P)$
$$
\hat{\Phi}(P') \equiv \sum_{Q',\overline{A}}
\Phi_{\overline{A}}(P',Q')\;|A_{1},...,A_{N}>
$$
where we have defined $P' \equiv \hat{P}Q \in \fS_{N}$, $Q' \equiv 
\hat{P}^{-1} \in \fS_{N}$ and the sum is over all types of 
excitations and all
corresponding sectors. The vector $|A_{1},...,A_{N}>$
belongs to $V_{1} \otimes ... \otimes V_{N}$, where 
$V= \spn\{2\uparrow,3\uparrow,4\uparrow,2\downarrow \}$ and 
represents one type of $N$
excitations. 

Thus, the relations between $\Phi(\hat{P} Q, \hat{P}^{-1})$ can be 
expressed using $S$-matrix presentation
$$
\begin{array}{l}
\hat{\Phi}(\Pi_{12}P) = S^{(1)}_{12}(\lambda_{1}-\lambda_{2}) 
\hat{\Phi}(P)
\end{array}
$$
where $S^{(1)}_{12}(\lambda_{1}-\lambda_{2})$  is
\bea
&& S^{(1)}_{12}(\lambda_{1}-\lambda_{2})
=  \label{def: S12 gl22gl2}
 \\
&&\left(%
\begin{array}{cccc|cccc|cccc|cccc}
-1& & & & & & & & & & & & & & & \\
& 1& & &  & & & &  & & & & & & & \\
& & t_{12}& & &  & & & r_{12}& & & & & & & \\
& & & t_{12}& & & & & & & & & r_{12} & & & \\
 \hline
& & & & 1 & & & & & & & & & & & \\
& & & &  & -1 & & &  & & & &  & & & \\
& & & &  &    & e^{-ik_{1}} & & &  & & &  & & & \\
& & & &  &    &   & e^{-ik_{1}} & & & & &  &  & & \\
 \hline
& & r_{12}& & &  & & & t_{12}& & & & & & & \\
& & & &  &    &  & & & e^{ik_{2}} & & &  & & & \\
& & & &  &    &    & & &   & -1 & & & & & \\
& & & &  &    &    & & &   &  & 0 &  & & -1 & \\
 \hline
& & & r_{12}& & & & & & & & & t_{12} & & & \\
& & & &  &    &   &  & & & & & &e^{ik_{2}} & & \\
& & & &  &    &    & & &   &  &-1 &  & & 0 & \\
& & & &  &    &    & & &   &   & &  & & & -1 
\end{array}%
\right)
% \begin{pmatrix}
%  |2\downarrow, 2\downarrow> \\
%  |2\downarrow, 2\uparrow> \\
%  |2\downarrow, 3\uparrow> \\
%  |2\downarrow, 4\uparrow> \\
%  \hline
%  |2\uparrow, 2\downarrow> \\
%  |2\uparrow, 2\uparrow> \\
%  |2\uparrow, 3\uparrow> \\
%  |2\uparrow, 4\uparrow> \\
%  \hline
%  |3\uparrow, 2\downarrow> \\
%  |3\uparrow, 2\uparrow> \\
%  |3\uparrow, 3\uparrow> \\
%  |3\uparrow, 4\uparrow> \\
%  \hline
%  |4\uparrow, 2\downarrow> \\
%  |4\uparrow, 2\uparrow> \\
%  |4\uparrow, 3\uparrow> \\
%  |4\uparrow, 4\uparrow> 
%   \end{pmatrix}
\,.\nonumber
\eea
Expressions for $t_{12}$ and $r_{12}$ were given in previous sections, 
for example in (\ref{r,t gl21gl2}).

The periodic boundary conditions can be written using the $S$-matrix and 
the vector $\hat{\Phi}(P)$
\begin{equation}
\hat{\Phi}(PC) = e^{ik_{p_{N}}L}\hat{\Phi}(P) \label{periodicity 
gl22gl2}
\end{equation}
and if we choose $P=C^{N-j}$, we arrive to the first auxiliary 
problem. Thus, for $j=1,...,N$,
\begin{equation}
S^{(1)}_{j+1 j}...S^{(1)}_{N j} S^{(1)}_{1 j}...S^{(1)}_{j-1 j}
\hat{\Phi}(id) = e^{ik_{j}L} \hat{\Phi}(id) \label{AuxPr1 gl22gl2}
\end{equation}
still with the convention $S^{(1)}_{a b} \equiv 
S^{(1)}_{a b}(\lambda_{a}-\lambda_{b})$. 

\medskip

The eigenvectors for this auxiliary problem are
given by
\begin{equation}
\phi[\bar{A}] = \sum_{\mathbf{x} \in [1,N]}
\Psi[\mathbf{x},\bar{A}]
e^{A_{1}}_{x_{1}}...e^{A_{N_{1}+N_{2}+N_{3}}}_{x_{N_{1}+N_{2}+N_{3}}}
\label{excited st aux pr1 gl22gl2}
\end{equation} 
with $A_{i} =(2,\uparrow);(3,\uparrow);(4,\uparrow)$ on the 
vacuum state filled by $e^{2 \downarrow}$ particles. We recall that
$N_{1}$ counts $e^{2 \uparrow}$ particles and $N_{2},N_{3}$ 
correspondingly $e^{3 \uparrow}$ and $e^{4 \uparrow}$ particles.

Comparing with the previous cases, the sector with two types of 
excitations $e^{3 \uparrow}$ and $e^{4 \uparrow}$  have been
already treated but we have an 
additional $\pi$-particle $e^{2 \uparrow}$. We write the eigenvector of this 
excitation similarly to (\ref{def: f func}) in  section 
\ref{sec:gl21-BAlevel2}:
\begin{equation}
\phi[2 \uparrow] = \sum_{x=1}^{N}
h_{x}(a) e^{2 \uparrow}_{x} = e^{2 \uparrow}_{a},\;\;\text{with}\;\; 
h_{x}(a)=\delta(x-a) \,.
\end{equation}
This form of eigenfunction is supported by the fact that the 
eigenfunction in (\ref{AuxPr1 gl22gl2}) should be independent of 
index $j$. The ansatz for general case with for all types of 
excitations $e^{3 \uparrow},e^{4 \uparrow}$ and $e^{2 \downarrow}$ 
can be written as, for 
$x_{q(1)}<x_{q(2)}<...<x_{q(N_{1}+N_{2}+N_{3})}$ and $Q \in 
\fS_{N_{1}+N_{2}+N_{3}}$
\begin{equation}
\Psi_{Q}(\mathbf{x},\bar{A}) = \sum_{P \in \cP_{f}} \Phi^{P Q}_{P^{-1}} 
\prod_{i=1}^{N_{1}} h_{x_{i}}(a_{n}) \prod_{n=1}^{N_{2}+N_{3}} 
f_{x_{n}}(a_{p(N_{1}+n)}) 
\label{ansatz aux pr1 gl22gl2}
\end{equation} 
where $\cP_{f}$ is the set of permutations acting on $\bar\pi$-particles only,
and $f_{x}(a)$ is defined in (\ref{f glNM}).  

The Bethe parameters $a_{1},...,a_{N_{1}}$ being the arguments of 
$h_{x}(a)$ are already quantized on the small chain $[1,N]$ and we 
choose them as:
$$
1 \leq a_{1}<...<a_{N_{1}} \leq N\,.
$$  

There are two different cases possible: $1)$ there exists a Bethe root 
$a_{\alpha} = j$ for some $\alpha \in [1,N_{1}]$, with $j$ being the 
index in (\ref{AuxPr1 gl22gl2}) and $2)$ there is no such Bethe 
root. In the first case, functions (\ref{excited st aux pr1 gl22gl2}) 
are eigenvectors of (\ref{AuxPr1 gl22gl2}) with the eigenvalue 
\begin{equation}
S^{(1)}_{j+1 j}...S^{(1)}_{N j} S^{(1)}_{1 j}...S^{(1)}_{j-1 j}
\hat{\Phi}(id) = (-1)^{N-N_{1}+1} e^{i k_{j} 
(N_{2}+N_{3})}\hat{\Phi}(id)\,.
\end{equation}

As we can see, the remaining parameters $a_{i}$ with $i \in 
[N_{1}+1,N_{1}+N_{2}+N_{3}]$ as well as the coefficients $\Phi^{P 
Q}_{P^{-1}}$ are not constrained in this first case: their Bethe 
equations  is obtained from the second case.

In the second case, when there is no Bethe root $a_{\alpha} = j$ for 
any $\alpha \in [1,N_{1}]$, (\ref{AuxPr1 gl22gl2}) implies the 
conditions on the coefficients $\Phi^{P Q}_{P^{-1}}$
\begin{equation}
\frac{\Phi^{P Q \Pi_{i i+1}}_{(P \Pi_{q(i) q(i+1)})^{-1}}}{\Phi^{P 
Q}_{P^{-1}}} = \frac{ ia_{p(q(i))} - ia_{p(q(i+1))} - 
\frac{u}{2}}{ia_{p(q(i))} - ia_{p(q(i+1))} + \frac{u}{2}},\;\;
\label{PhiPab=Phi pibar pibar}
\end{equation}
for all $p(q(i))$,$p(q(i+1))$ in $[N_{1}+1,N_{1}+N_{2}+N_{3}]$ and
\begin{equation}
\frac{\Phi^{P Q \Pi_{i i+1}}_{P^{-1}}}{\Phi^{P Q}_{P^{-1}}} = e^{-i 
k_{a_{p(q(i))}}} \sigma_{a_{p(q(i))}}(a_{p(q(i+1))})
\label{PhiPab=Phi pi pibar}
\end{equation}
for all $p(q(i)) \in [1,N_{1}]$, $p(q(i+1)) \in 
[N_{1}+1,N_{1}+N_{2}+N_{3}]$ and $\sigma_{i}(a)$ is defined in 
(\ref{sigma glNM}).

The calculations for the eigenvalue of (\ref{excited st aux pr1 
gl22gl2}) give
\begin{equation}
S^{(1)}_{j+1 j}...S^{(1)}_{N j} S^{(1)}_{1 j}...S^{(1)}_{j-1 j}
\hat{\Phi}(id) = (-1)^{N_{1}} \prod_{i=1}^{N_{2}+N_{3}} 
\sigma_{j}(a_{i+N_{1}}) \hat{\Phi}(id)\,.
\end{equation}
The periodic boundary condition on the coefficients $\Phi^{P 
Q}_{P^{-1}}$ is 
\beq
\frac{\Phi^{P Q C_{N_{1}+N_{2}+N_{3}}}_{P^{-1}}}{\Phi^{P Q}_{P^{-1}}} 
= (-1)^{N_{1}} \prod_{l=1}^{N} \sigma_{l}(a_{p(q(N_{1}+N_{2}+N_{3}))})
\label{eq:period-gl22}
\eeq
for $Q$ such that $q(N_{1}+N_{2}+N_{3}) \in 
[N_{1}+1,N_{1}+N_{2}+N_{3}]$. $C_{N_{1}+N_{2}+N_{3}}$ is a cyclic 
permutation, $C_{K} = \Pi^{1}_ {K}...\Pi^{K-1}_{K}$.

The difficulty in this model and of all models with $\pi$-particles is that
starting from the auxiliary problem we can not mix the Bethe roots of
different types of excitations. Indeed, in (\ref{ansatz aux pr1 gl22gl2})
the permutation acts only on $\bar\pi$-particles, while $\hat\Phi(P)$ mixes
\textit{a priori} any kind of particle. In this case we define again the
vector $\hat\Phi(P)$, but only for $\bar\pi$-particles as actually it was
defined in (\ref{def: hatPhi gl21gl2}):
\begin{equation}
\hat{\Phi}(P) \equiv \sum_{Q \in \fS_{N_{2}+N_{3}},\bar{A}} 
\Phi^{P}_{Q}(\bar{A})
|A_{q_{1}}, \ldots ,A_{q(N_{2}+N_{3})}>
\label{def: hatPhi gl22gl2}
\end{equation}
where the sum is over all types of $\bar\pi$ excitations and all
corresponding sectors. The vector $|A_{q_{1}} , \ldots ,A_{q_{K}}> $ 
is in $V_{1} \otimes  \ldots  \otimes V_{K}$ with 
$V=\spn\{e^{3\uparrow},e^{4\uparrow}\}$.

Working with the periodicity condition (\ref{eq:period-gl22}), 
we take for instance $Q=id$, which ensures that the constraint
 $q(N_{1}+N_{2}+N_{3}) \in 
[N_{1}+1,N_{1}+N_{2}+N_{3}]$ is satisfied. Thus, in the left hand 
side, in the coefficient $\Phi^{ 
\Pi^{p(1)}_{p(N_{1}+N_{2}+N_{3})}...\Pi^{p(N_{1}+N_{2}+N_{3}-1 
)}_{p(N_{1}+N_{2}+N_{3})} P}_{P^{-1}}$ we can identify which 
particles are permuted by $ \Pi^{p(k)}_{p(N_{1}+N_{2}+N_{3})}$ and we 
can apply corresponding relation (\ref{PhiPab=Phi pibar pibar}) or 
(\ref{PhiPab=Phi pi pibar}). Due to the fact that $P \in \cP_{f}$ does 
not act on $\pi$-particles, we can use relation (\ref{PhiPab=Phi pi pibar}) to
extract from the coefficient 
the action of the last 
$N_{1}$ permutations $\Pi^{p(j)}_{p(N_{1}+N_{2}+N_{3})}$, 
$j=1,\ldots,N_{1}$, and get:
\begin{equation}
\prod_{i=1}^{N_{1}} e^{-i k_{a_{i}}} 
\sigma_{a_{i}}(a_{p(N_{1}+N_{2}+N_{3})}) 
\Phi^{\Pi^{p(N_{1}+1)}_{p(N_{1}+N_{2}+N_{3})}...
\Pi^{p(N_{1}+N_{2}+N_{3}-1)}_{p(N_{1}+N_{2}+N_{3})} P}_{P^{-1}} 
=  (-1)^{N_{1}} \prod_{l=1}^{N} 
\sigma_{l}(a_{p(N_{1}+N_{2}+N_{3})}) \Phi^{P}_{P^{-1}}
\end{equation}
  
We can rewrite this condition using $\hat{\Phi}(P)$ notation. 
Choosing $P$ as a power of the cyclic permutation acting only on 
$\bar\pi$-particles, 
$$
P \equiv C^{N_{2}+N_{3} - m} = 
(\Pi^{N_{1}+1}_{N_{1}+N_{2}+N_{3}}...
\Pi^{N_{1}+N_{2}+N_{3}-1}_{N_{1}+N_{2}+N_{3}})^{N_{2}+N_{3} - m} \;,
$$ 
the calculation becomes equivalent to the one of section 
\ref{sec:gl21-BAlevel2}, and we obtain:
\begin{equation}
\prod_{i=1}^{N_{1}} e^{-i k_{a_{i}}} 
\sigma_{a_{i}}(a_{m}) \hat\Phi (C^{N_{2}+N_{3} - m 
+1 }) =  (-1)^{N_{1}} \prod_{l=1}^{N} 
\sigma_{l}(a_{m}) \hat\Phi(C^{N_{2}+N_{3} - m})
\,.
\end{equation}
Thus, for $m=N_{1}+1,...,N_{1}+N_{2}+N_{3}$
\bea
&&\prod_{i=1}^{N_{1}} e^{-i k_{a_{i}}} \sigma_{a_{i}}(a_{m})  
\prod_{\atopn{l=N_{1}+1}{l \neq m}}^{N_{1}+N_{2}+N_{3}} \left( \frac{ ia_{l} 
- ia_{m} - \frac{u}{2}}{ia_{l} - ia_{m} + \frac{u}{2}}\right) \left( 
P_{m\;m+1} ... P_{m\; N_{1}+N_{2}+N_{3}} P_{m\; 1}...P_{m\;m-1}  
\right) 
\hat\Phi (id) =
\nonu
&&= (-1)^{N_{1}} \prod_{l=1}^{N} \sigma_{l}(a_{m}) \hat\Phi (id)
\eea
and finally we get
\bea
&&\left( e^{2 \pi i\frac{\sum_{i=1}^{N_{3}}n_{i}}{N_{2}+N_{3}} }  
\prod_{i=1}^{N_{1}} e^{-i k_{a_{i}}} \prod_{\atopn{l=N_{1}+1}{l \neq 
m}}^{N_{1}+N_{2}+N_{3}} \left( \frac{ ia_{l} - ia_{m} - 
\frac{u}{2}}{ia_{l} - ia_{m} + \frac{u}{2}} \right)
- (-1)^{N_{1}} \prod_{\atopn{l=1}{l \not\in \bA}}^{N} 
\sigma_{l}(a_{m}) \right) \hat\Phi (id) = 0,
\nonu
&&m=N_{1}+1,...,N_{1}+N_{2}+N_{3}
\eea
where we have introduced the set 
$\bA=\{a_{1},a_{2},\ldots,a_{N_{1}}\}\subset [1,N]$ and the Bethe 
parameters $n_{j}$ that label the eigenfunctions of the cyclic 
permutation (as in section \ref{sec:glnm-BAlevel3}).

To stress the difference between the quantized parameters $a_{j}$, $j\leq 
N_{1}$, (that are similar to the parameters $n_{j}$), 
and the parameters $a_{j}$, $j>N_{1}$, we denoted the latter $b_{j}\equiv 
a_{j+N_{1}}$ in the Bethe equations written in section 
\ref{sec:Bethepi}.

\section{$gl(2|1)\oplus gl(2)$ and $gl(2|2)\oplus gl(2)$ Hamiltonians\label{sec:JordWig}}

In previous sections we considered examples of Hubbard model with algebra
$gl(2|1)_{\uparrow} \oplus gl(2)_{\downarrow}$ and its generalization to
$gl(\fn|\fm)_{\uparrow} \oplus gl(2)_{\downarrow}$ model. However only
examples with "small" algebras like $gl(2|1)_{\uparrow} \oplus
gl(2)_{\downarrow}$ or $gl(2|2)_{\uparrow} \oplus gl(2)_{\downarrow}$ seem
to find applications in physics. Performing different Jordan--Wigner
transformations one can write the corresponding Hamiltonians in terms of
creation and annihilation operators.

\subsection{Jordan--Wigner transformation}
We briefly recall some relations of Jordan--Wigner transformation 
\cite{JW} (for more detailed explanations, see e.g. \cite{XX}). 
The Jordan--Wigner transformation essentially consists in the 
construction of a mapping 
$$
E^{ij} \in gl(2^{p-1}|2^{p-1})
\leftrightarrow \{ c^{\dagger},c \; ; \; d^{\dagger},d \; ; \; 
e^{\dagger},e \; ; \; ... \},
$$
where $c,d,e,...$ are  fermionic operators. To present this 
construction, it is convenient to
introduce a matrix $X$
\begin{equation}
X=
\begin{pmatrix}
1 - n^{c} & c \\
c^{\dagger} & n^{c} 
\end{pmatrix}, \; \; n^{c}=c^{\dagger}c\,.
\end{equation}

Its entries $X_{\mu \nu}$  ($\mu,\nu=1,2$) have a natural grading given by
$[\mu ] + [\nu]$ where $[1] = 1$ and $[2] = 0$. The mapping 
is given by the relation
\begin{equation}
E^{ij} \leftrightarrow (-1)^{s} X^{(1)}_{\mu_{1} \nu_{1}} \otimes
X^{(2)}_{\mu_{2} \nu_{2}} \otimes ... \otimes X^{(p)}_{\mu_{p}
\nu_{p}}
\end{equation}
where to every value $i$ and $j$ $\in [1,2^{p}]$ is associated with an
element $\{ \mu_{1}, \; \mu_{2}, \; ... \;, \mu_{p} \}$ and 
$\{\nu_{1}, \; \nu_{2}, \; ... \;, \nu_{p} \}$ respectively with
$\mu_{i},\nu_{i}=1$ or $2$. Total grading is given by $s =
\sum_{i=2}^{p}[\mu_{i}] \left( \sum^{i-1}_{j=1}
([\mu_{j}]+[\nu_{j}]) \right)$.

Transformation for $gl(2)_{\downarrow}$ algebra is simply given by
identification of matrices $E^{ij}$ and $X_{\mu \nu}$. But since
$gl(2)$ algebra contain only bosonic operators, in order to satisfy 
the anticommutation relation between fermionic operators on different 
sites ($\{ c^{\dagger}_{a} , c_{b} \} = 0$), 
one should introduce some factor to
$E^{ij}_{a}$ (elementary matrix $E^{ij}$ on the site $a$). To present 
the result, we gather the different matrices $E^{ij}$, $i,j=1,2$ into 
a formal matrix $\EE$. Then, the Jordan--Wigner transformation for 
$gl(2)$ algebra reads
\begin{equation}
\EE_{a } =
\begin{pmatrix}
E^{11}_{a} & E^{12}_{a} \\
E^{21}_{a} & E^{22}_{a} \end{pmatrix}
\equiv
\begin{pmatrix}
1 - n^{c}_{a \downarrow} & c_{a \downarrow} \\
c^{\dagger}_{a \downarrow} & n^{c}_{a \downarrow} \end{pmatrix}
\prod_{x=a+1}^{L}(1 - 2 n^{c}_{x \downarrow} )\,.
\end{equation}

Now consider an example for the cases of $gl(2|2)_{\uparrow}$ and
$gl(2|1)_{\uparrow}$ algebras. For both algebra we perform one mapping and
in the case of smaller algebra ($gl(2|1)_{\uparrow}$) we remove a subspace.
We take the mapping $\{ 1 \rightarrow 11, \; 2 \rightarrow 22, \; 3
\rightarrow 12, \; 4 \rightarrow 21 \}$ as an example. In the initial
problem, elements of $gl(2|2)_{\uparrow}\oplus gl(2)_{\downarrow}$ and
$gl(2|1)_{\uparrow}\oplus gl(2)_{\downarrow}$ algebras transform into
fermionic operators which are considered to anticommute even for different
spins, e.g. $\{c^{\dagger}_{\downarrow} , c_{\uparrow} \} = 0$. This
implies that the total Jordan--Wigner transformation is written as
\begin{equation}
\EE_{a _{\uparrow}} \equiv
\begin{pmatrix}
(1 - n^{c}_{a \uparrow})(1 - n^{d}_{a \uparrow}) & -c_{a \uparrow}d_{a
\uparrow} & (1 - n^{c}_{a \uparrow})d_{a \uparrow} & -c_{a 
\uparrow}(1 -
n^{d}_{a \uparrow}) \\
c^{\dagger}_{a \uparrow}d^{\dagger}_{a \uparrow} & n^{c}_{a 
\uparrow}n^{d}_{a \uparrow}
& c^{\dagger}_{a \uparrow}n^{d}_{a \uparrow} & n^{c}_{a 
\uparrow}d^{\dagger}_{a
\uparrow} \\
(1 - n^{c}_{a \uparrow})d^{\dagger}_{a \uparrow} & n^{d}_{a 
\uparrow}c_{a
\uparrow} & (1 - n^{c}_{a \uparrow})n^{d}_{a \uparrow} & c_{a
\uparrow}d^{\dagger}_{a \uparrow} \\
-c^{\dagger}_{a \uparrow}(1 - n^{d}_{a \uparrow}) & n^{c}_{a 
\uparrow}d_{a
\uparrow} & d_{a \uparrow}c^{\dagger}_{a \uparrow} & n^{c}_{a 
\uparrow}(1 -
n^{d}_{a \uparrow}) \end{pmatrix} \prod_{x=1}^{L}(1 - 2 n^{c}_{x
\downarrow}),
\end{equation}
for $\EE=(E^{ij})_{i,j=1,\ldots,4} \in gl(2|2)$. 

For $\EE=(E^{ij})_{i,j=1,2,3}\in gl(2|1)$, it is given by 
the same matrix without line and row $3$
\begin{equation}
\EE_{a \uparrow}\equiv 
\begin{pmatrix}
(1 - n^{c}_{a \uparrow})(1 - n^{d}_{a \uparrow}) & -c_{a \uparrow}d_{a
\uparrow} & -c_{a \uparrow}(1 - n^{d}_{a \uparrow}) \\
c^{\dagger}_{a \uparrow}d^{\dagger}_{a \uparrow} & n^{c}_{a 
\uparrow}n^{d}_{a \uparrow}
& n^{c}_{a \uparrow}d^{\dagger}_{a \uparrow} \\
-c^{\dagger}_{a \uparrow}(1 - n^{d}_{a \uparrow}) & n^{c}_{a 
\uparrow}d_{a
\uparrow} & n^{c}_{a \uparrow}(1 - n^{d}_{a \uparrow}) \end{pmatrix}
\prod_{x=1}^{L}(1 - 2 n^{c}_{x \downarrow})
\end{equation}
where $n^{b}_{x \alpha} = b^{\dagger}_{x \; \alpha }\, b_{x \; 
\alpha}$ is the
particle number operator for $b = c,d$, and $\alpha = \uparrow, \;
\downarrow$. We have also standard relations between the operators:
\beq
\{ c^{\dagger}_{\alpha \; x}\,,\, c_{\beta \; y} \} = \delta_{xy} \,
\delta_{\alpha \beta}
\quad;\quad
\{ d^{\dagger}_{\alpha \; x}\,,\, d_{\beta \; y} \} = \delta_{xy} \,
\delta_{\alpha \beta}
\quad;\quad
\{ c_{\alpha \; x}\,,\, d_{\beta \; y} \} = 0\,,\quad 
\alpha,\beta = \uparrow,\downarrow\,. 
\eeq
Note that the operators $d^\dagger_{\downarrow}$, $d_{\downarrow}$ 
are 
not present in the construction, so that one can drop the arrow on 
the 
operators $d^\dagger_{\uparrow}$, $d_{\uparrow}$.

One can remark that the different choices of mappings $(i,j)$ on $\{
\bar\mu, \bar\nu \}$ is equivalent to some transformations on fermionic
operators' level (e.g. $c^{\dagger} \rightarrow c$, etc\ldots), therefore 
all the
Hamiltonians are equivalent in this sense and differs one from 
another by
changing the representation (e.g. from electrons to holes).

\subsection{$gl(2|2)\oplus gl(2)$ model.} 
The Hamiltonian of the model is given by 
\bea
H_{gl(2|2) \oplus gl(2)} &=& H_{Hub} + \sum_{x=1}^L(c^{\dagger}_{\uparrow 
\;
x+1 }c_{\uparrow \; x} + c^{\dagger}_{\uparrow \; x}c_{\uparrow \;
x+1})( n^{d}_{x \uparrow} n^{d}_{x+1 \uparrow} - n^{d}_{x
\uparrow} - n^{d}_{x+1 \uparrow} )  \nonu
&& + \sum_{x=1}^L\Big(c^{\dagger}_{\uparrow \; x+1
}c_{\uparrow \; x}d^{\dagger}_{\uparrow \; x+1 }d_{\uparrow \; x} +
c^{\dagger}_{\uparrow \; x}c_{\uparrow \; x+1}d^{\dagger}_{\uparrow 
\; x
}d_{\uparrow \; x+1}\Big) \nonu
&&  - 2 u \sum_{x=1}^L (1- 2n^{c}_{x \; \downarrow})
( 1- n^{c}_{x \; \uparrow } ) n^{d}_{x \; \uparrow} \nonu
&&  + \sum_{x=1}^L(d^{\dagger}_{\uparrow \; x+1
}d_{\uparrow \; x} + d^{\dagger}_{\uparrow \; x}d_{\uparrow \; x+1})(
n^{c}_{x \uparrow} n^{c}_{x+1 \uparrow} - n^{c}_{x \uparrow} -
n^{c}_{x+1 \uparrow} )
\eea
with
\begin{equation}
H_{Hub}= \sum_{\atopn{x=1}{ \alpha = \uparrow, \downarrow}}^L [ c^{\dagger}_{\alpha
\; x+1 }c_{\alpha \; x} + c^{\dagger}_{\alpha \; x}c_{\alpha \; x+1} 
] +
u \sum_{x=1}^L (1- 2n^{c}_{x \; \downarrow}) (1-2n^{c}_{x \; 
\uparrow})\,.
\label{eq:H-hub}
\end{equation}

The eigenfunctions for this Hamiltonian are made of creator operators 
$c^{\dagger}_{\uparrow},d^{\dagger}_{\uparrow}$ and 
$c^{\dagger}_{\downarrow}$.
They can be written in the following form and correspond to the 
solutions found in the previous sections:
\bea
\Phi^{(n)}_{N_{1},N_{2},N_{3}} &=& \sum_{\atopn{\vec{z}}{z_{k} \neq 
z_{l}}}
\sum_{\atopn{\vec{y}}{y_{k} \neq y_{l}}}
\sum_{\atopn{\vec{x}} {x_{i} \neq x_{j}, y_{k}}} 
\Psi'(\vec{x},\vec y,\vec z) 
\prod_{j=1}^{N_{1}-n} c^{\dagger}_{x_{j} \uparrow} \ 
\prod_{j=N_{1}+1}^{N_{1}+N_{2}-n} d^{\dagger}_{x_{j} \uparrow} 
\prod_{k=1}^{n} c^{\dagger}_{y_{k} \uparrow} d^{\dagger}_{y_{k} 
\uparrow} 
\prod_{j=1}^{N_{3}} c^{\dagger}_{z_{j} \downarrow} |0> 
\nonu
&=& (-1)^{N_{1}+N_{3}} 
\phi[\bar{A}]\,,\mb{with} n=0,...,\min (N_{1},N_{2})\\
\mbox{and} && \bar{A}
=(\overbrace{4\uparrow,...,4\uparrow}^{N_{1}-n},
\overbrace{3\uparrow,...,3\uparrow}^{N_{2}-n},
\overbrace{2\uparrow,...,2\uparrow}^{n},
\overbrace{2\downarrow,...,2\downarrow}^{N_{3}})\,.
\eea
$\phi[\bar{A}]$ is the eigenfunction given in (\ref{excited st 
glNM}). Remark that, in addition to the particles 
$c^{\dagger}_{\uparrow}$, $d^{\dagger}_{\uparrow}$ and 
$c^{\dagger}_{\downarrow}$ (corresponding to $4\uparrow$, $3\uparrow$ 
and $2\downarrow$ resp.), we have a doublet 
$c^{\dagger}_{\uparrow}d^{\dagger}_{\uparrow}$ corresponding to 
$2\uparrow$. The particles $c^{\dagger}_{\uparrow}$ and 
$c^{\dagger}_{\downarrow}$ can be identified with a (spin up and 
down) electron, while $d^{\dagger}_{\uparrow}$ can be viewed as a 
spin 0 fermion that can form bound state with the spin up electron.

The energy of the excited state $\Phi^{(n)}_{N_{1},N_{2},N_{3}}$
 reads
$$
E^{(n)}= L-2(N_{1}+N_{2}+N_{3}-n)+ 2 
\sum_{l=1}^{N_{1}+N_{2}+N_{3}-n} \cos k_{l}\,,
$$
where the parameters $k_{l}$ are Bethe roots defined by equations 
given in section 
\ref{sec:glnm-BAresu} (with $\fn=\fm=2$).

\subsection{$gl(2|1)\oplus gl(2)$ model.} 
The Hamiltonian of the model is given by 
\bea
H_{gl(2|1) \oplus gl(2)} &=& H_{Hub} + \sum_{x=1}^L(c^{\dagger}_{\uparrow 
\; x+1}
c_{\uparrow \; x} + c^{\dagger}_{\uparrow \; x}c_{\uparrow \; x+1})(
n^{d}_{\uparrow \; x} n^{d}_{x+1 \uparrow} - n^{d}_{\uparrow \; x} -
n^{d}_{\uparrow \; x+1} )  \nonu
&&  + \sum_{x=1}^L \Big(c^{\dagger}_{\uparrow \; x+1 }c_{\uparrow \;
x}d^{\dagger}_{\uparrow \; x+1 }d_{\uparrow \; x} + 
c^{\dagger}_{\uparrow \;
x}c_{\uparrow \; x+1}d^{\dagger}_{\uparrow \; x
}d_{\uparrow \; x+1}\Big)  \nonu
&&  - u \sum_{x=1}^L (1- 2n^{c}_{x \; \downarrow}) (
1- n^{c}_{x \; \uparrow } ) n^{d}_{\uparrow \; x}
\eea
where $H_{Hub}$ has been given in (\ref{eq:H-hub}).
Again, the eigenfunctions for this Hamiltonian correspond to the 
solutions found in the previous sections.
They have the form (for $N_{2}\leq N_{1}$)
\bea
\Phi_{N_{1},N_{2},N_{3}} &=& \sum_{\atopn{\vec{z}}{z_{k} \neq z_{l}}}
\sum_{\atopn{\vec{y}}{y_{k} \neq y_{l}}}
\sum_{\atopn{\vec{x}} {x_{i} \neq x_{j}, y_{k}}} 
\Psi(\vec{x},\vec y,\vec z) \prod_{l=1}^{N_{1}} c^{\dagger}_{x_{l} 
\uparrow} 
\prod_{l=1}^{N_{2}} c^{\dagger}_{y_{l} \uparrow} d^{\dagger}_{y_{l} 
\uparrow} \prod_{l=1}^{N_{3}} c^{\dagger}_{z_{l} \downarrow} 
|0> \nonu
&=& (-1)^{N_{1}+N_{3}} 
\phi[\bar{A}]\,,\mb{with}\bar{A}
=(\overbrace{3\uparrow,...,3\uparrow}^{N_{1}},
\overbrace{2\uparrow,...,2\uparrow}^{N_{2}},
\overbrace{2\downarrow,...,2\downarrow}^{N_{3}})
\eea
 $\phi[\bar{A}]$ has been defined in 
(\ref{eq:excited_st_gl21gl2}) and the corresponding eigenvalue reads
$$
E = L-2(N_{1}+N_{2}+N_{3})+ 2 \sum_{l=1}^{N_{1}+N_{2}+N_{3}} \cos 
k_{l}
$$
with Bethe roots $k_{l}$ obeying the equations given in section 
\ref{sec:gl21-BAresu}.

\section{Conclusion}

In this paper we presented the Bethe equations using the coordinate Bethe
ansatz for $gl(2|1)\oplus gl(2)$ and $gl(\fn|\fm)\oplus gl(2)$ generalized
Hubbard model. We wrote explicitly the Hamiltonians for several cases in
terms of fermionic creation and annhilation operators. Clearly, the full
derivation of the Bethe Ansatz Equations for generalized Hubbard models
\cite{XX,FFR2009} has to be accomplished: the case of $gl(\fn|\fm)\oplus
gl(3)$ and its generalizations to $gl(\fn|\fm)\oplus gl(\fn'|\fm')$
algebras are presently under investigation.

Applications to condensed matter physics deserve also some attention. The
models presented in section \ref{sec:JordWig} are generalization of Hubbard
models to several types of fermions. They could be of some relevance to
systems where electrons of different `types' or `colors' occur (for
instance on some ladder spin chain).

Although the link with AdS/CFT correspondence is not direct, the present
construction gives a way to introduce a phase in the Bethe equations of
Hubbard type models (see discussion in section \ref{sec:AdScft}). It is
thus worthwhile to look deeper at these models. They could be good
candidates for an integrable model close to the one underlying the SYM
theory, or give a new point of view for the wrapping problem.

Obviously, in all cases, the thermodynamical limit of these 
models needs also to be investigated.

% \section{biblio}

\end{document}